\documentclass[fleqn,10pt]{wlscirep}
\usepackage[utf8]{inputenc}
\usepackage[T1]{fontenc}
\title{Multi-critical topological transition at quantum criticality}

\author[1,2,$\dagger$]{Ranjith R Kumar}
\author[1,2,+]{Y R Kartik}
\author[1,2,+]{S Rahul}
\author[1,*]{Sujit Sarkar}
\affil[1]{Department of Theoretical Sciences, Poornaprajna Institute of Scientific Research, 4, Sadashivanagar, Bangalore-560 080, India.}
\affil[2]{Graduate Studies, Manipal Academy of Higher Education, Madhava Nagar, Manipal-576104, India.}

\affil[$\dagger$]{ranjith.btd6@gmail.com}
\affil[*]{sujit.tifr@gmail.com}

\affil[+]{these authors contributed equally to this work}


\begin{abstract}
The investigation and characterization of topological quantum 
phase transition between gapless phases is one of the recent 
interest of research in topological states of matter.
We consider transverse field Ising model with three spin interaction
in one dimension and observe a 
topological transition between gapless phases on one of the 
critical lines of this model.
We study the distinct nature of these gapless phases and 
show that they belong to different universality 
classes. The topological invariant number
	(winding number) characterize different topological phases for the
	different regime of parameter space. We observe the evidence of two multi-critical points, one is topologically
trivial and the other one is topologically active.
Topological quantum phase transition 
between the gapless phases on the critical line occurs 
through the non-trivial multi-critical point in the Lifshitz 
universality class. We calculate and analyze the behavior of 
Wannier state correlation function close to the multi-critical
point and confirm the topological transition between gapless phases. We show the breakdown of Lorentz 
invariance at this multi-critical point through the energy dispersion analysis. 
We also show that the scaling theories and curvature function renormalization group 
can also be effectively used to understand  the topological 
quantum phase transitions between gapless phases. The model Hamiltonian which we
	study is more applicable for the system with gapless excitations, where the conventional concept of topological quantum phase transition fails.
\end{abstract}
\begin{document}

\flushbottom
\maketitle

\thispagestyle{empty}

\section*{Introduction}

Quantum phase transitions is one of the fascinating subject in 
condensed matter physics. Landau's paradigm of spontaneous 
symmetry breaking describes continuous phase transitions 
successfully using local order parameter, which is 
finite at the ordered phase and vanishes at the critical point \cite{landau1937zh,miransky1994nuovo,sachdev2007quantum,vojta2003quantum}.
Contrary to this, topological quantum phase transitions (TQPT)--recently 
observed new class of phase transition--can be understood as a 
manifestation of topological properties of electronic band structure \cite{haldane1988model,kane2005quantum,kitaev2001unpaired}, 
instead of local order parameter. There is no spontaneous symmetry 
breaking associated, and hence it is not possible to define local order 
parameter for the transition between topologically distinct gapped phases. 
Topological gapped phases are distinguished by quantized topological 
invariants, which takes discrete values across TQPT points 
\cite{hasan2010colloquium,shen2012topological}.\\
Despite the failure of Landau's approach, recently, a theory of critical 
phenomena was found to be successful to  extract the critical behavior 
and obtain universality classes by identifying critical exponents using 
scaling relations in TQPTs 
\cite{griffith2018casimir,continentino2020finite,sun2017type,kempkes2016universalities,
	quelle2016thermodynamic}. These TQPT points are essentially quantum critical points (QCP), 
since they 
occur at zero temperature. One can define spacial and temporal 
characteristic lengths that have diverging behavior as we approach QCP. 
This diverging property of characteristic lengths with critical exponent 
$\nu$ (correlation length exponent) and $z$ (dynamical critical exponent), 
enable one to define universality classes of TQPTs 
\cite{continentino2017quantum,stanley1987introduction,continentino2017topological}. 
Localized edge modes in the topological non-trivial phases 
tend to delocalize and penetrate into the bulk as one approaches 
the TQPT point. The exponential decay of edge modes into the 
bulk depends on the distance to the topological transition ($g$) 
and is characterized by a length scale $\xi=|g|^{-\nu}$. This 
characteristic length $\xi$ can be referred as correlation length 
with critical exponent $\nu$ \cite{zhou2008finite,chu2009coherent}. 
Correlation length exponent can be obtained using several 
approaches including the numerical studies of penetration 
length of the edge modes as a function of the distance to 
the transition \cite{griffith2018casimir,continentino2020finite}, 
and also from the scaling properties of the Berry connection 
\cite{chen2017correlation,chen2018weakly,chen2019universality}. 
At QCP energy dispersion $E_k$ is found to be $E_k\propto k^z$, 
where $z$ is dynamical critical exponent. Expanding the energy 
dispersion around the QCP and identifying the dominant momentum 
one can find the value of $z$, which governs the shape of the 
spectra at the gap closing point \cite{rufo2019multicritical}.\\
As one approaches TQPT point the system exhibits scale invariance. 
Exploiting this property, a scaling theory, analogous to the 
Kadanoff's scaling theory of conventional critical phenomena \cite{kadanoff1966scaling}, 
has been proposed \cite{chen2017correlation,chen2016scaling}. 
The topological invariant--calculated by integrating curvature function 
over the whole Brillouin zone in the momentum space--takes integer 
values for topological gapped phases and changes abruptly at the 
critical point. The curvature function diverges at the critical point 
signaling the critical behavior of TQPT point. Based on this behavior 
of curvature function a renormalization group (RG) approach 
has been developed \cite{van2018renormalization,molignini2018universal}. 
A knot-tying scaling procedure is proposed based on the divergence in 
the curvature function at the critical point. This scaling procedure 
changes the curvature function and drives the system to its fixed point
configuration, without changing the topology of the band structure. 
Since the topological invariant does not change during this process, 
the RG flow lines distinguish between distinct topological gapped 
phases. In one dimensional systems this scaling procedure is 
analogous to stretching a string until the knots are revealed \cite{luo2019advanced}. 
This curvature function renormalization group (CRG) has been 
used in studying the topological phase transition in, Kitaev model, 
Su-Schrieffer-Heeger model \cite{chen2016scaling}, periodically
driven systems \cite{molignini2018universal,molignini2020generating}, 
systems without inversion symmetry \cite{abdulla2020curvature}, 
models with $Z_2$ invariant \cite{chen2016scaling2}, quantum walks that simulate one
and two-dimensional Dirac models \cite{panahiyan2020fidelity}, multi-critical 1D 
topological insulator \cite{malard2020scaling} and also in 
interacting systems \cite{chen2018weakly,kourtis2017weyl} etc. \\
All these characterizing tools mentioned above have been widely 
used to distinguish between gapped phases separated by a  topological transition. However, 
the appearance of 
transition between stable gapless phases with trivial and non-trivial 
topological characters have also been observed in a wide class of 
magnetic systems \cite{beri2010topologically,chen2013critical,brzezicki2017topological,
	liang2018intermediate,nasu2018successive}. Exponentially localized edge
modes at the QCPs, in one and two-dimensional symmetry protected topological phases, are 
stable to disorder and can give rise to topologically
distinct gapless phases \cite{gapless1,gapless2}.\\
\textbf{Motivation:} In this work, 
we are motivated to study the TQPT occurring between two gapless
phases through a Lorentz symmetry breaking point.  
We consider transfer field Ising model (TFIM) with three spin interaction \cite{kopp2005criticality}, 
where the study of edge modes at criticality has revealed the appearance and 
disappearance of localized edge modes at one of the quantum critical lines 
with corresponding change in the parameter values \cite{rahul2019topology}. 
In other words, both topological and non-topological characters appear on 
the same critical line for different parameter regimes. This provides an 
interesting platform to study TQPT between gapless phases as well as to 
understand the validity of characterizing tools in 
identifying this transition. \\
Motivation of this work is twofold. First 
is to prove that,
indeed the critical line possess distinct gapless phases and there is a 
TQPT between these phases occurring through a multi-critical point which breaks the Lorentz
invariance in our model Hamiltonian. Second one is to perform this using 
characterizing techniques that have been used to distinguish 
between gapped phases, thereby validating the reliability of 
these techniques to distinguish between gapless phases.
We also show the relation between the breaking of Lorentz invariance 
and topological quantum phase transition at the multi-critical point.
This 
phenomenon can be analogously understood from the topological semimetals, 
where the Dirac points confluence to form quadratic dispersion at a critical point which
breaks the Lorentz symmetry \cite{volovik2018exotic,soluyanov2015type}.\\
There are several studies on multi-critical behavior and 
topological transition using conventional RG techniques in the
literature \cite{PhysRevB.94.041101,PhysRevB.93.235112,PhysRevLett.114.185701,sarkar2016physics,sarkar2020study}. 
The conventional RG captures the 
physics of correlated topological systems with local Coulomb interaction
in one, two and three dimensions. However, here we adopt CRG based on the 
diverging behavior of curvature function as we approach the topological quantum 
critical point. Since the curvature function encapsulates the topological signatures
of the band structure, its prominent behavior near the transition point is 
promising and sufficient to address the unconventional topological 
transition between gapless phases in our model.

\section*{Model Hamiltonian and Topological Quantum Phase Diagram}\label{model}
\begin{figure}
	\includegraphics[width=11cm,height=7cm]{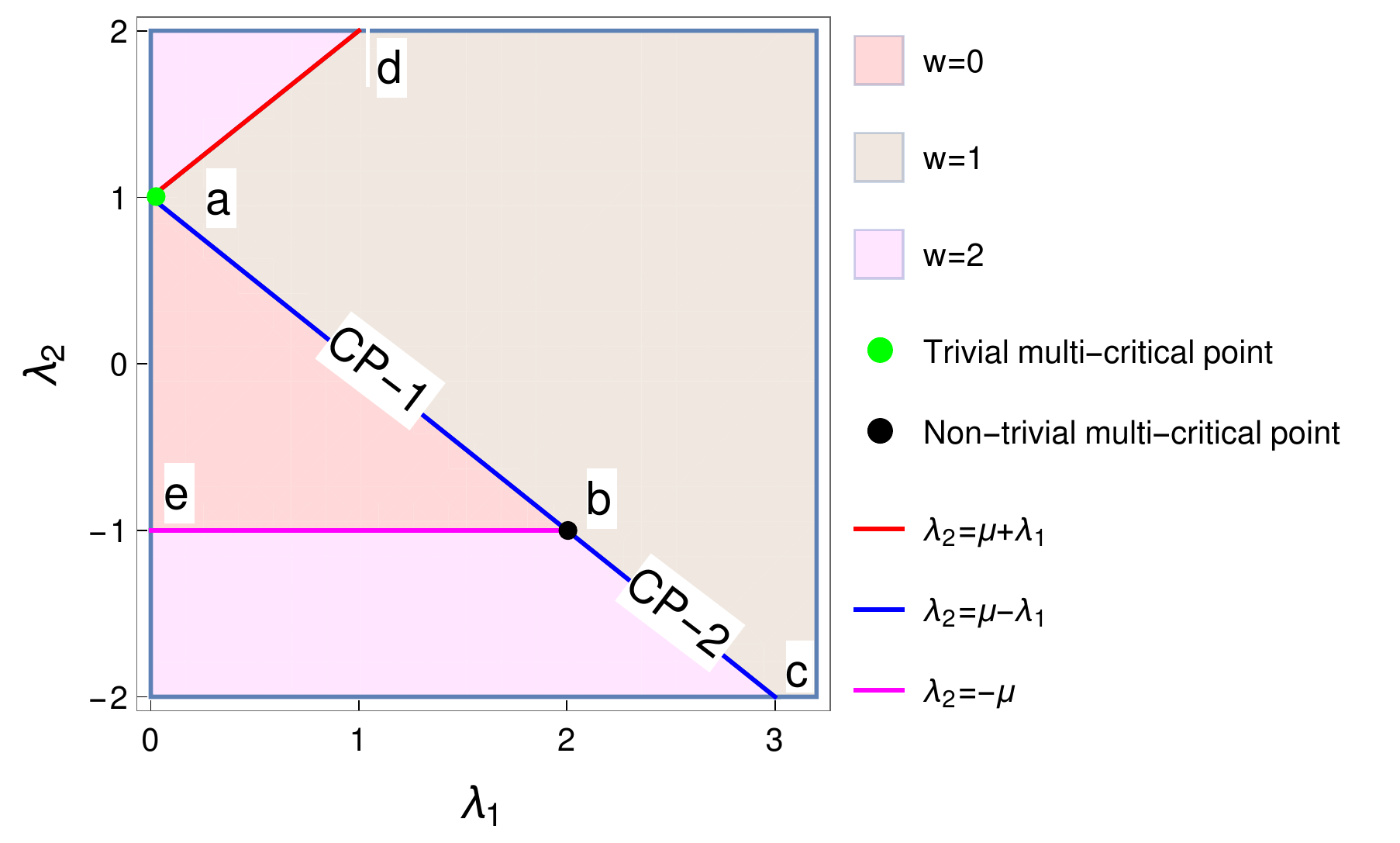}
	\caption{Topological phase diagram of model Hamiltonian for $\mu=1$. Line `ac' 
		represents the critical line $\lambda_2=\mu-\lambda_1$ (blue line), line `be' represents
		the critical line $\lambda_2=-\mu$ (magenta line) and line `ad' represents the critical line 
		$\lambda_2=\mu+\lambda_1$ (red line). Points `a' and `b' are multi-critical points (green and black dots respectively) which 
		differentiate between three distinct gapped phases with $w=0,1,2$ (represented in different colors). Here CP-1 
		is critical/gapless phase  for the transition between $w=0$ and $w=1$. CP-2 is 
		critical/gapless phase for the transition between $w=1$ and $w=2$.} \label{phase_diag2}
\end{figure}

We consider transverse field Ising model with three spin interaction \cite{kopp2005criticality,sarkar2018quantization}
\begin{equation}
H=-\sum_{i} \left(\lambda_1 \sigma_{i}^z \sigma_{i-1}^z + \lambda_2 \sigma_{i}^x \sigma_{i-1}^z \sigma_{i+1}^z + \mu \sigma_{i}^x\right) ,
\end{equation}
where $\sigma^{x,z}$ are Pauli matrices. Performing Jordan-Wigner transformation
$\sigma_{i}^x=1-2c_i^{\dagger}c_i$ and $\sigma_{i}^z= - \prod_{j<i} (1-2c_{j}^{\dagger}c_{j}) (c_{i} + c_{i}^{\dagger})$, the model Hamiltonian  can be written in spinless fermionic form as
\begin{equation}
H = -\mu \sum_{i=1}^{N} (1 - 2 c_{i}^{\dagger}c_{i}) - \lambda_1 \sum_{i=1}^{N-1} (c_{i}^{\dagger}c_{i+1} + c_{i}^{\dagger}c_{i+1}^{\dagger} + h.c) - \lambda_2 \sum_{i=2}^{N-1} ( c_{i-1}^{\dagger}c_{i+1} +  c_{i+1} c_{i-1} + h.c),
\label{jw} 
\end{equation}
where nearest neighbor superconducting gap is equal to nearest neighbor 
hopping amplitude $\lambda_1$ and next nearest neighbor superconducting 
gap is equal to next nearest neighbor hopping amplitude $\lambda_2$.
In this equation, $c_i^{\dagger}(c_i)$ is creation (annihilation) fermionic operator 
and $h.c$ represents the Hermitian conjugate. It is a one-dimensional mean-field 
	model for a triplet superconductor. The three spin interaction added to the transverse 
	field Ising model can be physically realized in realistic Hamiltonians since the term 
	is generated through real-space renormalization group treatments \cite{kopp2005criticality}. \\
The 
Bloch Hamiltonian of Eq.\ref{jw}, which is a $2\times2$ matrix, can be written as 
\begin{equation}
\mathcal{H}(k) =  \chi_{z} (k) \sigma_z - \chi_{y} (k) \sigma_y ,
\label{APS}
\end{equation}
where $ \chi_{z} (k) = -2 \lambda_1 \cos k - 2 \lambda_2 \cos 2k 
+ 2\mu,$ and $ \chi_{y} (k) = 2 \lambda_1 \sin k + 2 \lambda_2 \sin 2k.$ 
The excitation spectra can be obtained as 
\begin{equation}
E_k=\pm\sqrt{\chi_{z}^2 (k) + \chi_{y}^2 (k)}.
\end{equation}
This model supports topological distinct gapped phases (i.e $w=0,1,2$) separated 
by the three quantum critical lines as shown in Fig.\ref{phase_diag2}. The energy 
gap closes at these quantum critical lines, $\lambda_2=\mu+\lambda_1$, $\lambda_2=\mu-\lambda_1$ and $\lambda_2=-\mu$, 
obtained for momentum $k_0=\pm\pi$, $k_0=0$ and $k_0=\cos^{-1}(-\lambda_1/2\lambda_2)$ respectively. 
The topological angle can be written as $\phi_k=\tan^{-1}\left(\chi_{y} (k)/\chi_{z} (k)  \right) $.\\
The 
	model has been studied previously in different contexts \cite{kopp2005criticality,niu2012majorana,sarkar2018quantization,rahul2019topology}.
	The model was first introduced by the authors of Ref. \cite{kopp2005criticality}
	to study the persistence of quantum criticality at high temperature in correlated systems. 
	The authors of Ref.\cite{niu2012majorana} has studied the physics 
	of Majorana zero modes in the gapped phases of this model with both broken and 
	unbroken time-reversal symmetry. 
	One of the authors (S.S) has studied the quantization of geometric phase with integer and fractional 
	topological characterization for this model in Ref.\cite{sarkar2018quantization}. 
	Very recently authors of Ref.\cite{rahul2019topology} have solved the problem of bulk-boundary correspondence
	at the quantum critical lines and discussed the principle of least topological invariant 
	number at the criticality.\\
In this work we intent to show explicitly that there exist a TQPT 
between two gapless phases (CP-1 and CP-2 in Fig.\ref{phase_diag2}) on the critical line $\lambda_2=\mu-\lambda_1$ 
through a multi-critical point $\lambda_1=2\mu$ (point `b' in Fig.\ref{phase_diag2}).  We also explore the 
nature of transition and critical behavior implementing the scaling law 
of critical theories and show that these characterizing tools, which are 
used to characterize the transition between gapped phases, are also 
efficient tools to characterize the TQPT between gapless phases.\\
There are two multi-critical points at the intersections of the critical lines. 
For the parameter value $\mu=1$ a multi-critical point with an emergent 
$U(1)$ symmetry exist at $(\lambda_1,\lambda_2)=(0,1)$ \cite{niu2012majorana}.
This multi-critical point `a' in the phase diagram (Fig.\ref{phase_diag2}) occurs at the 
intersection of the critical lines $\lambda_2=\mu+\lambda_1$ and $\lambda_2=\mu-\lambda_1$.
It posses linear spectra at the gap closing momenta $k=0$ and $k=\pm \pi$
and does not break the Lorentz 
invariant. Since it does not involve any topological transition between gapless phases 
on a critical line, we consider it a trivial multi-critical point.
Another multi-critical point exist at $(\lambda_1,\lambda_2)=(2,-1)$. 
This multi-critical point `b' in the phase diagram 
occurs at the intersection of critical lines $\lambda_2=\mu-\lambda_1$ and $\lambda_2=-\mu$. 
Since it posses quadratic spectra at $k=0$ and breaks Lorentz invariance, we consider it to be a non-trivial 
multi-critical point. This is 
exactly the point $\lambda_1=2\mu$, through which TQPT between gapless 
phases occur.\\
\begin{figure}
	\includegraphics[width=10.5cm,height=3cm]{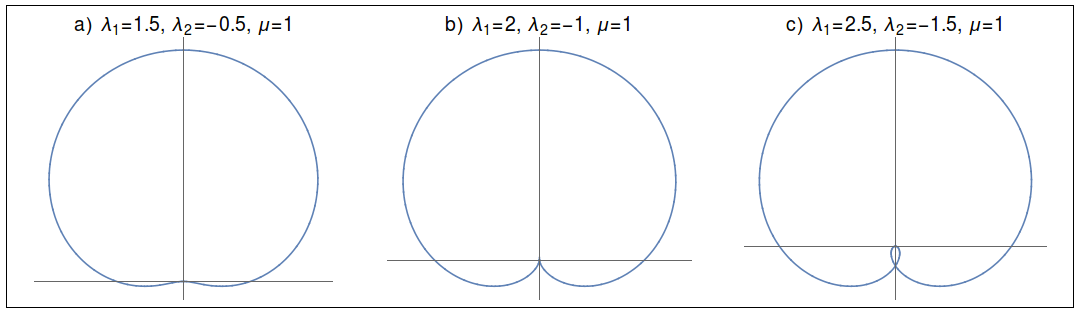}
	\caption{Parameter space for pseudo spin-vector on the critical line $\lambda_2=\mu-\lambda_1$. 
		a) trivial gapless phase b) multi-critical point c) non-trivial gapless phase.}
	\label{auxilary}
\end{figure}
The transition can be verified by investigating behavior 
of pseudo spin-vector in the parameter space \cite{zhang2015topological,sarkar2018quantization}. 
The model Hamiltonian can be expressed in terms of 
pseudo spin-vector as
\begin{equation}
\mathcal{H}(k)= \boldsymbol{\chi}(k).\boldsymbol{\sigma}
\end{equation} 
where $ \chi_{z} (k) = -2 \lambda_1 \cos k - 2 \lambda_2 \cos 2k 
+ 2\mu,$ and $ \chi_{y} (k) = 2 \lambda_1 \sin k + 2 \lambda_2 \sin 2k$. The pseudo 
spin-vector takes a closed curve in the parameter space 
around the origin for a set of parameter values representing a gapped 
phase. For gapless phase the curve passes through the origin and this 
behavior is characteristic of criticality. In Fig.\ref{auxilary} we have 
shown the behavior of pseudo spin-vector in the parameter space on 
the critical line $\lambda_2=\mu-\lambda_1$. The curve is always 
closed and passes through the origin indicating the criticality. As one 
goes from Fig.\ref{auxilary}.(a) to (c), system is passing 
from topologically trivial gapless phase to non-trivial gapless phase 
through a multi-critical point (Fig.\ref{auxilary}.(b)). Trivial gapless 
phase is the phase boundary between $w=0$ and $w=1$ gapped 
phases, as well as, non-trivial gapless phase is the phase boundary 
between $w=1$ and $w=2$ gapped phases. The non-trivial gapless 
phase is characterized by the emergence of secondary loop which 
passes through the origin. Therefore this behavior of pseudo spin-vector 
suggest that there exist a TQPT between two gapless phases on the
critical line $\lambda_2=\mu-\lambda_1$.

\section*{Results and discussion}

\subsection*{Energy dispersion and critical exponents }\label{energy_dis_criti_expo}
\begin{figure}[t]
	\includegraphics[width=14.8cm,height=8cm]{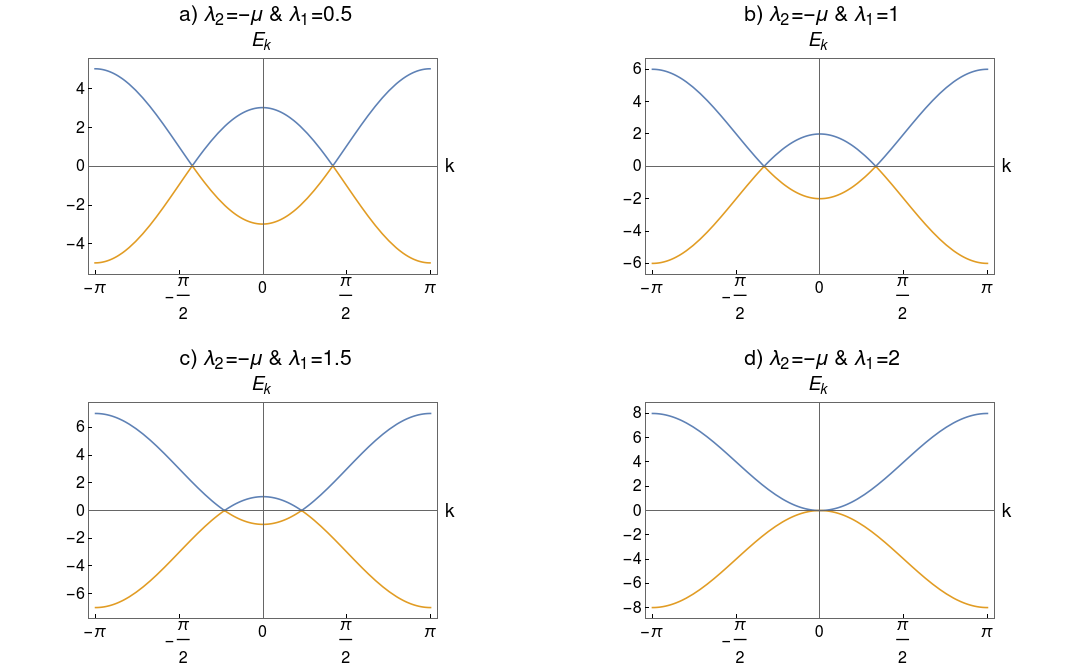}
	\caption{(a,b,c) Spectra on the critical line $\lambda_2=-\mu$ (with $\mu$=1). 
		There are two gapless points around which the spectra is 
		linear (i.e., $E_k\propto k$) which implies $z=1$. (d) Spectra
		at multi-critical point with $\lambda_1=2\mu$ and $\lambda_2=-\mu$. 
		Two gapless points confluence at $k=0$ where the spectra is quadratic 
		(i.e., $E_k\propto k^2$) and $z=2$. }
	\label{spectra}
\end{figure}

One can distinguish between the universality classes of the gapless 
phases by calculating the values of critical exponents. In this section 
we calculate the correlation length critical exponent ($\nu$) and dynamical 
critical exponent ($z$) for the two gapless phases on the critical line 
$\lambda_2=\mu-\lambda_1$.\\
The spectra of this model on the critical line $\lambda_2=\mu-\lambda_1$ 
is gapless and linear  for $\lambda_1<2\mu$, and quadratic 
for $\lambda_1\geq2\mu$. 
On the critical line $\lambda_2=-\mu$ spectra has two gapless points at 
the two incommensurate momenta, $\pm k_0$, symmetric about the point 
$k=0$ as shown in Fig.\ref{spectra}(a-c). As we approach multi-critical 
point on this critical line, the two incommensurate points confluence at 
$k_0=0$ (i.e, $(\lambda_1,\lambda_2)=(2,-1)$), as shown in Fig.\ref{spectra}(d). 
Therefore the spectra is non-relativistic (breaks Lorentz invariance) and 
become quadratic in nature instead of linear.
Energy dispersion for one dimensional system close quantum critical point 
can be written as $E_k=\sqrt{|g|^{2\nu z}+k^{2z}}$, where $\nu$ is 
correlation length critical exponent and $z$ is dynamical critical exponent \cite{rufo2019multicritical}. 
At the critical point the gap function $\Delta=|g|^{2\nu z}$ 
should go to zero, therefore $E\propto k^z$.\\
The energy dispersion expanded around the gap closing momenta $k_0 = 0$ 
can be written as
\begin{equation}
E_k=\pm\sqrt{(2\mu-2\lambda_1-2\lambda_2)^2 + C_2 k^2 +C_4 k^4} , \label{dispersion}
\end{equation}
where $C_2= (16\lambda_2\mu+4\lambda_1\mu-4\lambda_1\lambda_2)$ and 
$C_4= \frac{1}{3}(\lambda_1\lambda_2-\lambda_1\mu-16\lambda_2\mu)$. 
Gap function $g^{2\nu z}=(2\mu-2\lambda_1-2\lambda_2)^2$ 
implies $\nu z =1$. At QCP the gap function goes to zero and the 
shape of the spectra can be obtained as $E_k \propto k^z$, 
by identifying the dominant coefficient among $C_2$ and $C_4$. 
Above the multi-critical point (trivial gapless phase, i.e., $\lambda_1<2\mu$) 
one can observe that the coefficient of quadratic term $C_2$ is 
much larger than $C_4$. Therefore quadratic term dominate implying 
$E_k\propto k$, hence $z=1$. Similarly below the multi-critical point  
(non-trivial gapless phase, i.e., $\lambda_1>2\mu$) one can find that 
$C_4$ dominates over $C_2$ and the spectra $E_k \propto k^2$ 
implying the value of $z=2$. At the multi-critical point 
(i.e, $\lambda_1=2\mu$ and $\lambda_2=-\mu$) the 
coefficient $C_2=0$, which entails $z=2$ since $E_k\propto k^2$. 
Therefore the dynamical critical exponent is found to have $z=1$ 
with linear spectra at the trivial gapless phase and $z=2$ with 
quadratic spectra at transition point (multi-critical point)  as well 
as non-trivial gapless phase.
Once the dynamical critical exponent $z$ is obtained one can also 
obtain the value of correlation length critical exponent $\nu$ from 
the condition $\nu z=1$ in our model. Thus in the trivial gapless 
phase the critical exponents are $z=1$ and $\nu=1$ and in the 
non-trivial gapless phase $z=2$ and $\nu=\frac{1}{2}$. Note that 
the situation $C_2=C_4$ is not possible on the critical line since it requires 
$\lambda_1$ to be complex. Equating $C_2$ and $C_4$ results in 
$\lambda_1\propto \left(4 \mu -i \sqrt{177} \mu \right)$, which is not
possible in our model, implying $C_2\neq C_4$.\\
This observation suggest that these two gapless phases belong to 
different universality classes since their critical exponents has 
different set of values. This entails the fact that there is a TQPT in 
the Lifshitz universality class with $z=2$ and $\nu=\frac{1}{2}$ \cite{volovik2017lifshitz,leite2004new,rufo2019multicritical}, 
between two distinct gapless phases through multi-critical 
point. Thus in this study the breaking of Lorenz invariance occurs at the
Lifshitz universality class.\\
\begin{figure}
	\includegraphics[width=8cm,height=3.5cm]{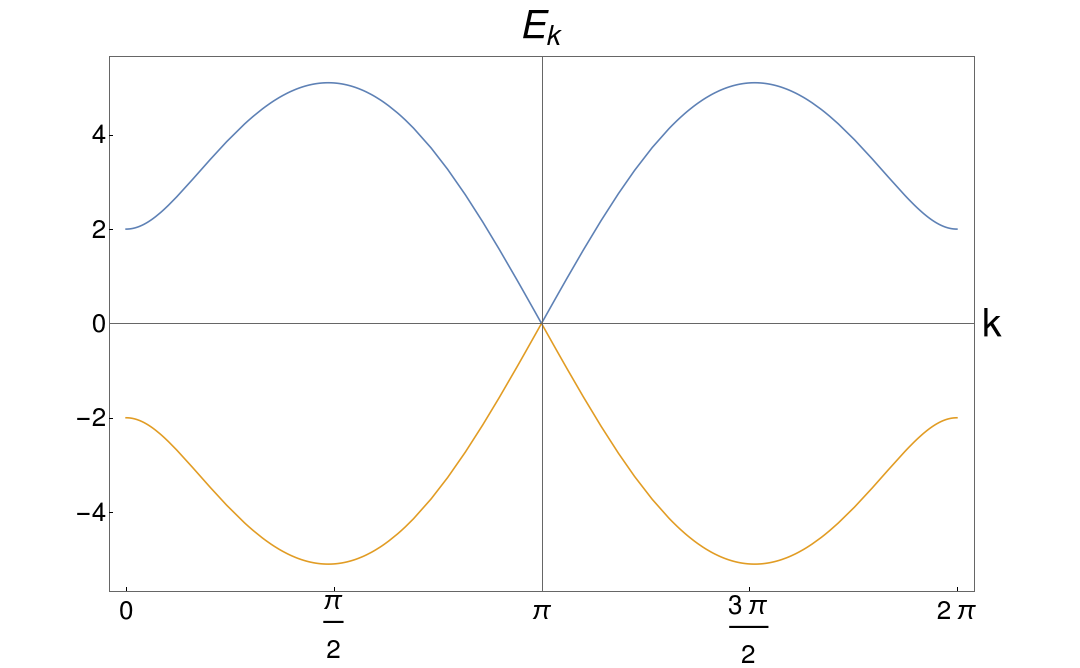}
	\caption{Spectra on the critical line $\lambda_2=\mu+\lambda_1$}
	\label{spectra_kpi}
\end{figure}
For completeness we also calculate the critical exponents for the critical theory at 
$\lambda_2=\mu+\lambda_1$. The spectra on this line is found to 
be linear in $k$ as shown in Fig.\ref{spectra_kpi}, which 
implies the value of dynamical critical exponent to be $z=1$. 
Spectra close to $k_0=\pm\pi$ can be written as 
\begin{equation}
E_k = \pm \sqrt{(2\mu+2\lambda_1-2\lambda_2)^2+C_2 k^2 + C_4 k^4},
\end{equation}
where $C_2=(4\lambda_1\lambda_2-4\lambda_1\mu+4\lambda_2\mu)$ and $C_4=\frac{1}{3}(\lambda_1\mu-\lambda_1\lambda_2-16\lambda_2\mu)$. 
At the QCP gap function goes to zero and coefficient $C_2$ dominates 
over $C_4$, implying $E_k \propto k$. Therefore the spectra at the gap 
closing point is linear and dynamical critical exponent $z=1$. The gap 
function $g^{2\nu z}= (2\mu+2\lambda_1-2\lambda_2)^2$ implies 
$\nu=1$.\\
We have shown the breakdown
	of Lorentz invariant symmetry at the multi-critical point. The
	authors of Ref. \cite{roy2018global,sur2019unifying,tarruell2012creating} have shown explicitly that the break 
	down of Lorentz invariance also occur for graphene and 3D Weyl semimetal. The
	authors of Ref. \cite{roy2018quantum} have shown explicitly the transformation from
	the Dirac semimetal to band insulator QCP at $\Delta =0 $, ($\Delta $ is the energy
	scale), where the quasiparticle spectra is two momentum space dimension. 
	In $x$-direction, it is linear in $k$ and in the $y$-direction it is quadratic ($k^2$).
	But the model Hamiltonian which we have studied is one dimension, therefore only
	one component has appeared.\\
We confirm the results for our model by calculating the critical exponents 
from the Berry connection approach and also show the presence of 
TQPT between gapless phases using CRG analysis in the next section. \\

\subsection*{Curvature function renormalization group}\label{CRG-general}
At first, we briefly review the curvature function renormalization group (CRG) 
method which encapsulates the critical behavior of a system during 
topological phase transition. Let us consider a system with a set of 
parameters $\mathbf{M}=(M_1,M_2,M_3,...)$, which upon tuning 
appropriately changes the underlying topology of the system and 
induces topological phase transition. The curvature function 
$F(k,\mathbf{M})$ at momentum $k$ dictate the topological properties of the system. 
Integral of this curvature 
function over a Brillouin zone defines topological invariant number 
which characterizes a gapped phase. For 1D systems it reads
\begin{equation}
w=\int\limits_{-\pi}^{\pi}\frac{dk}{2\pi} F(k,\mathbf{M}).
\end{equation}
Change in this topological invariant number involves the phase transition 
between the distinct gapped phases. For 1D systems Berry 
connection is the curvature function. Since Berry connection is 
gauge dependent, one can choose the gauge for which $F(k,\mathbf{M})$ 
can be written in Ornstein-Zernike form around the high symmetry 
point (HSP) $k_0$,\\
\begin{equation}
F(k_0+\delta k,\mathbf{M})=\frac{F(k_0,\mathbf{M})}{1\pm\xi^2\delta k^2}, \label{Lorenzian}
\end{equation}
where $\delta k$ is small deviation from HSP, and $\xi$ is characteristic 
length scale. As the system approaches critical point to undergo 
topological phase transition i.e, $\mathbf{M}\rightarrow \mathbf{M}_c$, 
curvature function diverges and changes sign as system moves across 
critical point
\begin{equation}
\lim_{\mathbf{M}\rightarrow \mathbf{M}_c^+}F(k_0,\mathbf{M})= -\lim_{\mathbf{M}\rightarrow \mathbf{M}_c^-}F(k_0,\mathbf{M})=\pm \infty. \label{div-curvature}
\end{equation}
Based on the divergence of the curvature function near HSPs, a scaling 
theory has been developed. For given $\mathbf{M}$ we find new 
$\mathbf{M^{\prime}}$ which satisfies
\begin{equation}
F(k_0,\mathbf{M^{\prime}})=F(k_0+\delta k,\mathbf{M}), \label{scaling}
\end{equation}
where $\delta k$ satisfies $F(k_0+\delta k,\mathbf{M})=F(k_0-\delta k,\mathbf{M})$. 
If the topology of the system at $\mathbf{M}$ and at fixed point $\mathbf{M}_f$ 
are same then the curvature function can be written as 
$F(k,\mathbf{M})=F_f(k,\mathbf{M}_f)+F_d(k,\mathbf{M}_d)$, 
where  $F_f(k,\mathbf{M}_f)$ is curvature function at fixed point 
and $F_d(k,\mathbf{M}_d)$ is deviation from the fixed point. 
Applying Eq.\ref{scaling} iteratively makes  $F_d(k,\mathbf{M}_d) \rightarrow 0$, 
implying gradual decrease in the deviation of curvature function 
from the fixed point configuration. Hence $F(k,\mathbf{M})\rightarrow F_f(k,\mathbf{M}_f)$. 
Finding the map from $\mathbf{M}$ to $\mathbf{M^{\prime}}$ 
iteratively, broadens the curvature function $F(k_0, \mathbf{M})$ 
until it reaches fixed point. This iterative procedure yields RG flow 
in parameter space indicating critical points of the system.
Generic RG equation of parameters $\mathbf{M}$ can be obtained 
by expanding Eq.\ref{scaling} to leading order and writing 
$d\mathbf{M}=\mathbf{M^{\prime}}-\mathbf{M}$ and $\delta k^2=dl$, as \cite{chen2016scaling,van2018renormalization}
\begin{equation}
\frac{d\mathbf{M}}{dl}= \frac{1}{2} \frac{\partial_k^2 F(k,\mathbf{M})|_{k=k_0}}{\partial_{\mathbf{M}}F(k_0,\mathbf{M})}. \label{CRG_eq}
\end{equation}
The critical point can be defined by the condition $|\frac{d\mathbf{M}}{dl}|=\infty$, 
and fixed point can be defined by the condition $|\frac{d\mathbf{M}}{dl}|=0$. 
As we approach critical point, along with the divergence of the 
curvature function (Eq.\ref{div-curvature}), characteristic length 
$\xi$ in Eq.\ref{Lorenzian} also diverges
\begin{equation*}
\lim_{\mathbf{M}\rightarrow \mathbf{M}_c}\xi=\infty.
\end{equation*}
These divergences in $F(k_0,\mathbf{M})$ and $\xi$ give rise to 
divergent behavior characterized by the critical exponents
\begin{equation}
F(k_0,\mathbf{M}) \propto |\mathbf{M}-\mathbf{M}_c|^{-\gamma} \;\;\;,\;\;\;\;\;\; \xi \propto |\mathbf{M}-\mathbf{M}_c|^{-\nu}. \label{critical-exponents}
\end{equation}
In conventional Landau theory of phase transition with order 
parameter, correlation function plays prime role. The same can not be 
defined for topological phase transitions since there is no local order 
parameter. However, a correlation function in terms of a matrix 
element between Wannier states of distant home cells is proposed 
to characterize the topological phase transition \cite{chen2017correlation}. 
This Wannier state correlation function $\lambda_R$, can be obtained 
from Fourier transform of the curvature function for 1D systems as
\begin{equation}
\lambda_R = \int\limits \frac{dk}{2\pi} e^{ikR}F(k,\mathbf{M}). \label{correlation}
\end{equation}
Substituting the Ornstein-Zernike form of curvature function 
yields $\lambda_R\propto e^{-\frac{R}{\xi}}$. This suggest that $\xi$ can be 
treated as correlation length of topological phase transition with critical 
exponent $\nu$. Similarly curvature function at HSP, $F(k_0,\mathbf{M})$ 
has the notion of susceptibility in the Landau paradigm with the critical 
exponent $\gamma$. These critical exponents define the universality 
class of a model undergoing topological phase transition. A generic 
scaling law--imposed by the conservation of topological invariant--can be 
deduced for the critical exponents as
\begin{equation}
\gamma=\sum\limits_{i=1}^{D} \nu_i, \label{scaling_law}
\end{equation}
where $D$ is the dimensionality of the system. Thus for 1D systems we 
have $\gamma=\nu$ \cite{chen2017correlation}. The CRG method has been used to understand 
topological transition between gapped phases. Here we use this method to understand the topological transition 
between previously discussed gapless phases in our 
model. We calculate the RG equations and critical exponents for the 
critical theories between both gapped and gapless transitions and 
ensure the reliability of this method.
\subsubsection*{CRG for the transition between gapped phases}
\begin{figure}
	\includegraphics[width=13cm,height=7cm]{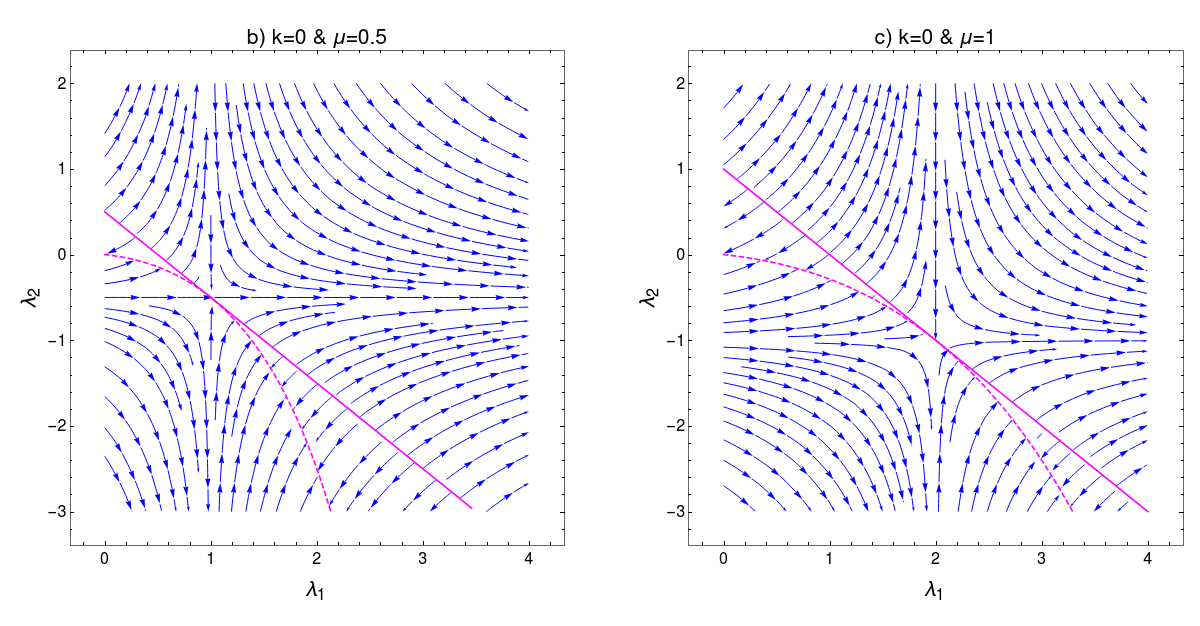}
	\caption{Flow diagram for $k_0=0$ in $\lambda_1$-$\lambda_2$ 
		plane for a) $\mu=0.5$ and b) $\mu=1$. The RG flow directions are 
		pointed by the arrows. The critical lines are shown as solid lines and fixed 
		lines as dashed lines. Analyzing RG flow, distinct topological phases and 
		the transition between them can be understood.} \label{CRG_flow_k0}
\end{figure}

\begin{figure}
	\includegraphics[width=13cm,height=7cm]{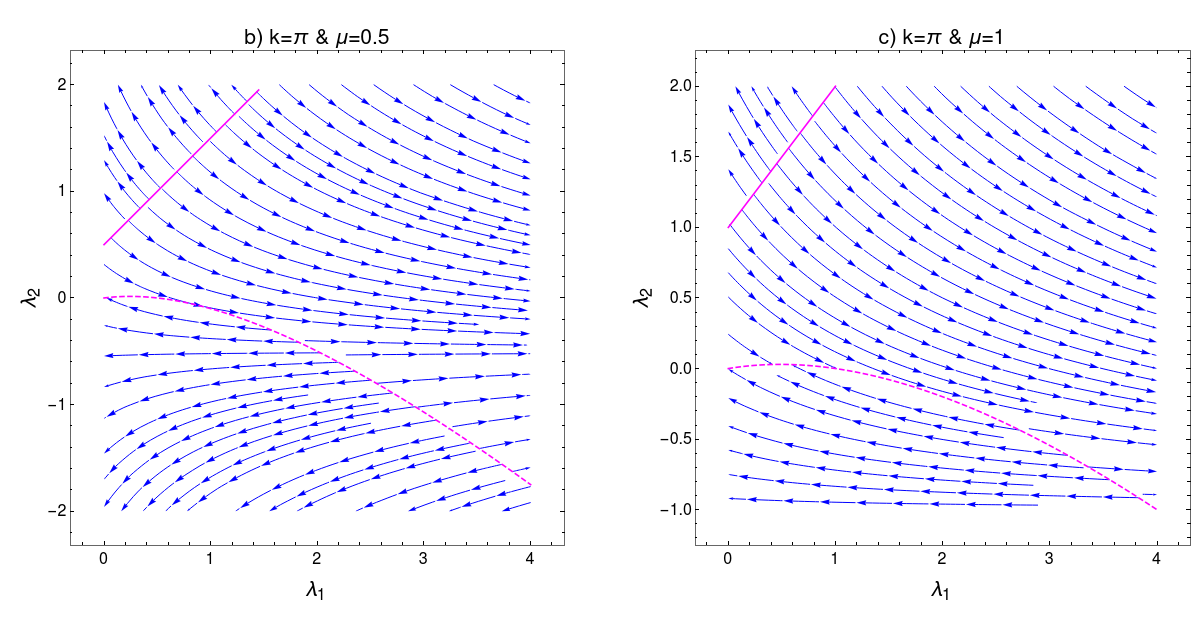}
	\caption{Flow diagram for $k=\pi$ in $\lambda_1$-$\lambda_2$ 
		plane for a) $\mu=0.5$ and b) $\mu=1$. The RG flow directions are 
		pointed by the arrows. The critical and fixed 
		lines are shown as solid and dashed lines respectively.} \label{CRG_flow_kpi}
\end{figure}

In this section we perform CRG for the topological transition across the 
critical line $\lambda_2=\mu-\lambda_1$, i.e, between the gapped 
phases with $w=0,1$ and $2$. The objective of this discussion is to 
distinguish between the distinct critical phases CP-1 and CP-2 . We derive RG equations to confirm the 
topological transition between the gapped phases (between $w=0,2$ and $w=1$). 
We derive critical exponents for the CP-1 and CP-2 through 
Berry connection approach \cite{rufo2019multicritical} to characterize 
their universality classes. Transition between the CP-1 and CP-2 through 
the multi-critical point `b' is studied in the next 
section.\\
The curvature function can be calculated as
\begin{equation}
\begin{split}
F(k,\mathbf{M})&= \dfrac{d\phi_k}{dk}\\
&= \dfrac{d}{dk}\left[ \tan ^{-1}\left(\frac{2 \lambda_2 \sin (2 k) +2 \lambda_1 \sin (k)}{2 \mu -2 \lambda_2 \cos (2 k) -2 \lambda_1 \cos (k)}\right)\right] \\
&= \frac{\lambda_1  \cos (k) (\mu -3 \lambda_2)+2 \lambda_2 \mu  \cos (2 k)-\lambda_1^2-2 \lambda_2^2}{2 \lambda_1 \cos (k) (\lambda_2-\mu )-2 \lambda_2 \mu  \cos (2 k)+\lambda_1^2+\lambda_2^2+\mu ^2}, \label{curvature-function}
\end{split}
\end{equation}
\begin{figure}
	\includegraphics[width=14cm,height=7cm]{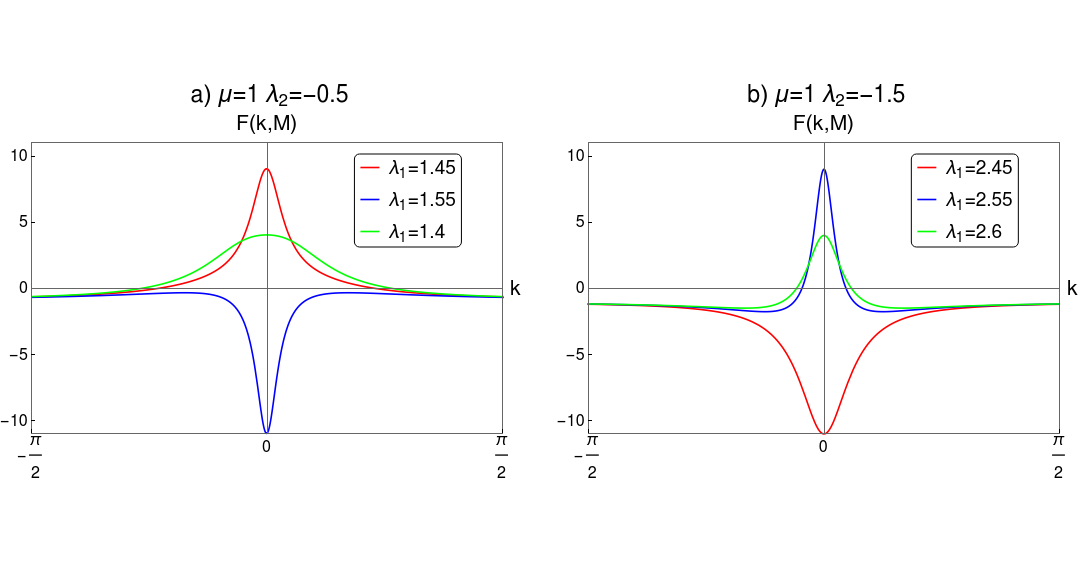}
	\caption{ Curvature function $F(k,\mathbf{M})$ near the HSP $k_0=0$ 
		plotted for $\lambda_2<0$. (a) Curvature function plotted for several 
		values of $\lambda_1$ at $\mu=1$ and $\lambda_2=-0.5$ for the transition 
		between $w=0$ and $w=1$. (b) Curvature function plotted for several 
		values of $\lambda_1$ at $\mu=1$ and $\lambda_2=-1.5$ for the transition 
		between $w=2$ and $w=1$. In both (a) and (b) the plot is around the QCPs, 
		which defines the topological transition between gapped phases. As the QCP is 
		approached, curvature function diverges at HSP and flips sign as we 
		cross it. The scaling procedure proposed in CRG will fit here since the 
		condition $F(k_0,\mathbf{M^{\prime}})=F(k_0+\delta k,\mathbf{M})$ 
		is satisfied.} \label{curvature_plot}
\end{figure}
where $\mathbf{M}= \left\lbrace \mu, \lambda_1, \lambda_2\right\rbrace $. 
Behavior of $F(k,\mathbf{M})$ near the QCPs for the transition between 
gapped phases is shown in Fig.\ref{curvature_plot}. The transition 
between $w=0$ and $w=1$ is shown in Fig.\ref{curvature_plot}(a) for the 
parameter values $\lambda_2=-0.5$ and $\mu=1$. For this transition 
critical point is obtained for $\lambda_1=1.5$ at $k_0=0$. In Fig.\ref{curvature_plot}(b), 
curvature function for transition between $w=2$ and $w=1$ for parameter 
values $\lambda_2=-1.5$ and $\mu=1$ is shown, where the critical point 
appear for $\lambda_1=2.5$ at $k_0=0$. Curvature function tend to 
diverge as we approach the critical points and flips sign as we cross it. This confirms that $F(k,\mathbf{M})$ 
takes the Ornstein-Zernike form of Eq.\ref{Lorenzian} around the 
HSP $k_0=0$.
RG flow equations can be constructed now to see the flow line's behavior 
in the parameter space to understand the topological transition in the 
model. The RG equations can be derived for $k_0=0$ as (refer to `Method' section
for a detailed derivation)
\begin{equation}
\dfrac{d\lambda_1}{dl}=\frac{\text{$\lambda_1 $}^2+\text{$\lambda_1$} (\mu -\text{$\lambda_2 $})+8 \text{$\lambda_2 $} \mu }{2 (\text{$\lambda_1 $}+\text{$\lambda_2 $}-\mu )}\label{RG-lam1},
\end{equation}
\begin{equation}
\dfrac{d\lambda_2}{dl}=-\frac{(\text{$\lambda_2 $}+\mu ) \left(\text{$\lambda_1 $}^2+\text{$\lambda_1 $} (\mu -\text{$\lambda_2 $})+8 \text{$\lambda_2 $} \mu \right)}{2 (\text{$\lambda_1 $}-2 \mu ) (\text{$\lambda_1 $}+\text{$\lambda_2 $}-\mu )} \label{RG-lam2},
\end{equation}
\begin{equation}
\dfrac{d\mu}{dl}=-\frac{(\text{$\lambda_2 $}+\mu ) \left(\text{$\lambda_1 $}^2+\text{$\lambda_1 $} (\mu -\text{$\lambda_2 $})+8 \text{$\lambda_2 $} \mu \right)}{2 (\text{$\lambda_1 $}+2 \text{$\lambda_2 $}) (\text{$\lambda_1 $}+\text{$\lambda_2 $}-\mu )} \label{RG-mu}.
\end{equation}  
For a constant 
value of $\mu$, Eq.\ref{RG-lam1} and Eq.\ref{RG-lam2} satisfy the conditions 
\begin{equation}
\left| \dfrac{d\lambda_1}{dl}\right| =\left| \dfrac{d\lambda_2}{dl}\right| = 
\infty \quad \text{and} \quad
\left| \dfrac{d\lambda_1}{dl}\right| =\left| \dfrac{d\lambda_2}{dl}\right| = 0.
\end{equation}
One can observe critical line and fixed line respectively at 
$\lambda_2=\mu-\lambda_1$ and $\lambda_2=\frac{\lambda_1(\lambda_1+\mu)}{\lambda_1-8\mu}$. 
RG flow lines for the coupling parameters $\lambda_1$ and $\lambda_2$ 
are depicted in Fig.\ref{CRG_flow_k0} for $k_0=0$.
It consists of two figures for different values of $\mu$. In each figure the quantum
critical line and fixed line are represented as solid and dashed lines respectively.  
Direction of the RG flow, 
in the $\lambda_1$-$\lambda_2$ plane, is shown by the arrows, which 
signals the presence of critical and fixed lines. The critical line is denoted 
by solid line in the flow diagram which traces a line $\lambda_2=\mu-\lambda_1$ 
as predicted analytically. This line distinguish between, $w=0$ and $w=1$ 
gapped phases for $\lambda_1<2\mu$ and $w=2$ and $w=1$ gapped phases 
for $\lambda_1>2\mu$ for $\mu\neq 0$. The RG flow of coupling parameters  $\lambda_1$ and 
$\lambda_2$ flows away from the critical line and towards the stable fixed 
line as shown in Fig.\ref{CRG_flow_k0}(a) and (b). One can dubiously 
distinguish between $w=0$ and $w=2$ gapped phases based on the flow lines, 
which flows towards $\lambda_1=2\mu$ in $w=2$ phase and towards the fixed line 
in $w=0$ phase. \\
Multi-critical point appear exactly at the intersection of
critical and fixed lines, i.e at the point $(\lambda_1,\lambda_2)=(2\mu,-\mu)$.
This intersection point can be obtained analytically 
by equating critical and fixed line equations, which yield a quadratic equation 
$\lambda_1^2-4\mu\lambda_1+4\mu^2=0$. The solution of this 
quadratic equation is $\lambda_1=2\mu$ which is the multi-critical point
for the HSP $k_0=0$. The curvature function is found to be diverging at 
this point. This multi-critical point distinguish the critical phases 
$\lambda_1<2\mu$ and $\lambda_1>2\mu$ 
on the critical line, whose physics can also 
be captured by the CRG method which is discussed in the next section.\\
In order to show the distinct nature of CP-1 and CP-2, 
we calculate the critical exponents, explained in Eq.\ref{critical-exponents}, 
and characterize their universality classes. Set of critical exponents 
$(z,\nu,\gamma)$ characterize the critical phases which governs the 
transition between $w=0$ and $w=1$ as well as $w=2$ and $w=1$ 
gapped phases. To calculate these critical exponents we first expand 
the Hamiltonian terms $\chi_z$ and $\chi_y$ from Eq.\ref{APS}, around 
the HSP $k_0=0$.
\begin{align}
\chi_z&= (2\mu-2\lambda_1-2\lambda_2)+\frac{(8\lambda_2+2\lambda_1)}{2} \delta k^2\\
\chi_y&= (4\lambda_2+2\lambda_1)\delta k,
\end{align}
where $(2\mu-2\lambda_1-2\lambda_2)=\delta g$, such that  $F(k_0,\delta g) = 
F_0|\delta g|^{-\gamma}$ and $\xi = \xi_0 |\delta g|^{-\nu}$. We substitute 
$B=\frac{(8\lambda_2+2\lambda_1)}{2}$ 
and $A=(4\lambda_2+2\lambda_1)$ and write the Berry connection in 
Ornstein-Zernike form in Eq.\ref{Lorenzian} as (refer to `Method' section for details)
\begin{equation}
F(k,\delta g)= \frac{\left( \frac{2BA\delta k^2-A(\delta g+B\delta k^2)}{\delta g^2}\right) }{1 + \frac{(2 \delta g B+A^2)}{\delta g^2} \delta k^2 + \frac{B^2}{\delta g^2}\delta k^4 }
= \frac{F(k_0,\delta g)}{1+\xi^2 \delta k^2+\xi^4\delta k^4}.\label{curvature1}
\end{equation}
here we observe that among the coefficients of $\delta k^2$, the second 
term diverges more quickly and becomes dominant as we approach QCP. 
For transition between gapped phases $w=0$ and $w=1$, coefficient 
$\delta k^2$ term dominates over the coefficient $\delta k^4$ term 
implying $\xi \propto |\delta g|^{-1}$, thus the correlation length and 
dynamical critical exponents $\nu=1$ and $z=1$ respectively. For 
transition between gapped phases $w=2$ and $w=1$, coefficient 
$\delta k^4$ term dominates over the coefficient $\delta k^2$ term 
implying $\xi \propto |\delta g|^{-\frac{1}{2}}$, thus the critical 
exponents can be obtained as $\nu=\frac{1}{2}$ and $z=2$. The 
curvature function at the HSP $k_0=0$ can be obtained 
as $ F(k_0,\delta g)= \frac{2(\lambda_1+2\lambda_2)}{\delta g}$. 
As we approach critical line  $\lambda_2=\mu-\lambda_1$ the 
curvature function $F(k_0,\delta g) \propto |\delta g|^{-1}$ 
implying the curvature function critical exponent to be $\gamma=1$.\\
Summarizing above results suggest that the set of critical 
exponents for CP-1 between 
$w=0 \quad \text{and} \quad w=1$ are $(\nu,z,\gamma)=(1,1,1)$ and
for CP-2 between $w=2 \quad \text{and} \quad w=1$ 
are $(\nu,z,\gamma)=(\frac{1}{2},2,2)$. This clearly indicate 
that the two gapless phases belong to different universality classes. 
There is a TQPT between these two gapless phases through 
multi-critical point which we discuss in the next section. 
This result coincide with the results that we obtained 
from energy dispersion analysis.\\
Note that for CP-1 the scaling law in Eq.\ref{scaling_law} is obeyed, 
while for CP-2 it is violated. 
The dynamical critical exponent is found to take the value $z=1$ 
for CP-1 since the spectra is linear in $k$ around the 
gap closing point. In 
the case of CP-2, the spectra is found to be quadratic 
in $k$ around the gap closing point which yields $z=2$. For this 
case one can write an effective form of Eq.\ref{curvature1} around the HSP as 
\begin{equation}
F(k,\delta g)=\frac{F(k_0,\delta g)}{(1+\xi^4\delta k^4)}.
\end{equation}
Integrating this over its 
width $\xi_i^{-1}$ for the conservation of topological invariant, 
yields the scaling law $\gamma=2\sum\limits_{i=1}^{D} \nu_i$. 
Thus when $z=2$ the scaling law will get modified into $\gamma=2\nu$ 
for 1D systems (refer to `Method' section for details). \\
\begin{figure}
	\includegraphics[width=6.5cm,height=4.5cm]{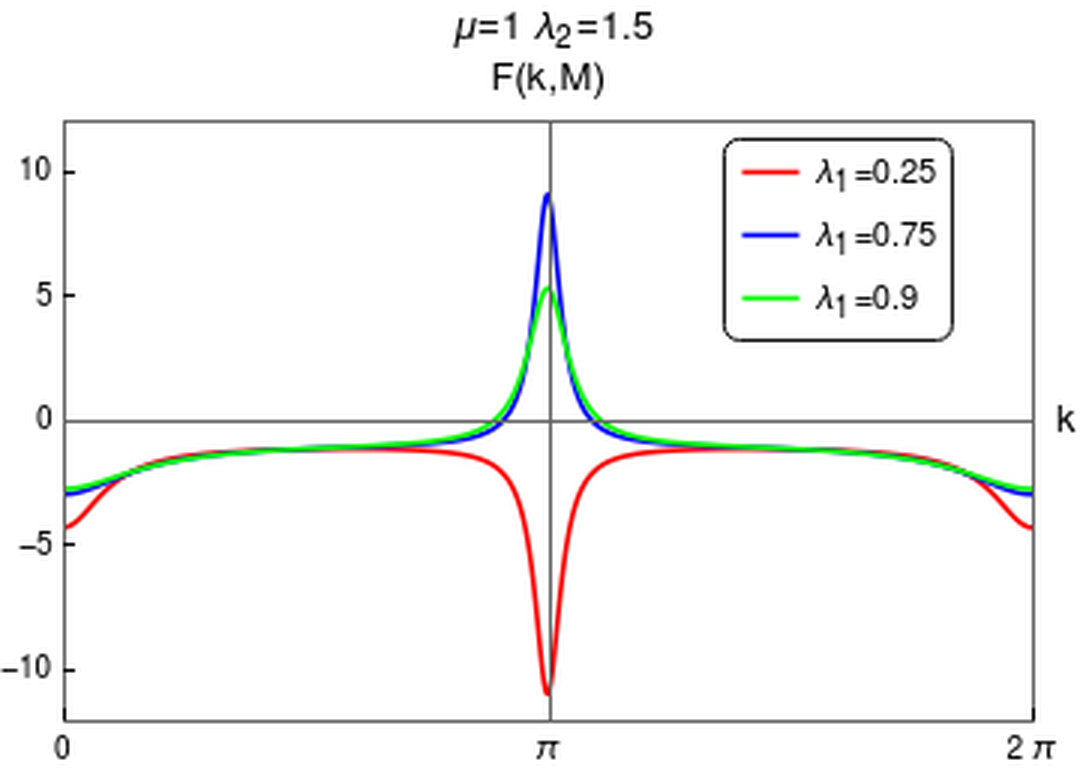}
	\caption{The behavior of the curvature function around the HSP 
		$k_0=\pi$ for $\lambda_2>0$. Several values of $\lambda_1$, around the critical value $\lambda_1=0.5$, are plotted at $\mu=1$ and $\lambda_2=1.5$. Curvature function shows suitable behavior to perform CRG as it diverges at HSP on approaching critical point.} \label{curvature_kpi}
\end{figure}
In order to verity this modification in scaling law, we perform the CRG 
for the HSP $k=\pi$ which address the topological transition between 
gapped phases $w=2$ and $w=1$ for $\lambda_2>0$. This transition 
happens through the critical line $\lambda_2=\mu+\lambda_1$.
As we approach this QCP the curvature function in Eq.\ref{curvature-function}, 
diverges at the HSP $k_0=\pi$ as shown in Fig.\ref{curvature_kpi}
and takes the Ornstein-Zernike form around this HSP. RG flow equations 
for the coupling parameters $\lambda_1$, $\lambda_2$ and $\mu$
can be derived as (refer to `Method' section for a detailed derivation)
\begin{equation}
\frac{d\lambda_1}{dl}=\frac{\text{$\lambda_1 $}^2+\text{$\lambda_1$} (\text{$\lambda_2 $}-\mu)+8 \text{$\lambda_2 $} \mu }{2 (\text{$\lambda_1 $}-\text{$\lambda_2 $}+\mu )}, \label{RGklam1}
\end{equation}
\begin{equation}
\frac{d\lambda_2}{dl}=-\frac{(\text{$\lambda_2 $}+\mu ) \left(\text{$\lambda_1 $}^2+\text{$\lambda_1 $} (\text{$\lambda_2 $}-\mu)+8 \text{$\lambda_2 $} \mu \right)}{2 (\text{$\lambda_1 $}+2 \mu ) (\text{$\lambda_1 $}-\text{$\lambda_2 $}+\mu )}, \label{RGklam2}
\end{equation}
\begin{equation}
\frac{d\mu}{dl}=-\frac{(\text{$\lambda_2 $}+\mu ) \left(\text{$\lambda_1 $}^2+\text{$\lambda_1 $} (\text{$\lambda_2 $}-\mu)+8 \text{$\lambda_2 $} \mu \right)}{2 (\text{$\lambda_1 $}-2 \text{$\lambda_2 $}) (\text{$\lambda_1 $}-\text{$\lambda_2 $}+\mu )}.
\end{equation}
For a constant value of $\mu$, Eq.\ref{RGklam1} and Eq.\ref{RGklam2} satisfy the conditions
\begin{equation}
\left| \dfrac{d\lambda_1}{dl}\right| =\left| \dfrac{d\lambda_2}{dl}\right| = \infty \quad \text{and} \quad
\left| \dfrac{d\lambda_1}{dl}\right| =\left| \dfrac{d\lambda_2}{dl}\right| = 0.
\end{equation} 
The critical line and fixed line can be found at $\lambda_2=\mu+\lambda_1$ and $\lambda_2=\frac{\lambda_1(\mu-\lambda_1)}{8\mu+\lambda_1} $
respectively.
The RG flow diagram for coupling parameters at $k_0=\pi$ is shown 
in Fig.\ref{CRG_flow_kpi}.
It consists of two figures for different values of $\mu$. In each figure the quantum
critical line and fixed line are represented as solid and dashed lines respectively.    
The critical line $\lambda_2=\mu+\lambda_1$, 
represented as solid line in the flow diagram, distinguish between 
$w=2$ and $w=1$ gapped phases. The RG flow lines flowing away 
from this critical line indicate the TQPT between these gapped phases. 
The fixed lines are represented as dashed curve in Fig.\ref{CRG_flow_kpi}(a) and (b). 
A part of this fixed line is stable where flow lines flows towards it and a 
part is unstable where flows are away from it for $\mu \neq 0$. The intersection of these 
critical and fixed lines can 
be obtained analytically by equating critical and fixed line equations. 
This yield $\lambda_1=-\mu$, which indicate there is no intersection point 
for positive $\mu$ or $\lambda_1$ values.\\
We verify the value of critical exponent $\nu$ using Berry 
connection approach. Expanding the Hamiltonian terms $\chi_z$ 
and $\chi_y$ of Eq.\ref{APS} around the HSP $k_0=\pi$ upto 
first order in $k$ and writing the Berry connection $F(k_0,\mathbf{M})$ 
in the form of Eq.\ref{Lorenzian} yields (refer to `Method' section for details)
\begin{equation}
F(k,\delta g)= \frac{\left( \frac{A}{\delta g}\right) }{1 + \frac{(A^2)}{\delta g^2} \delta k^2}
= \frac{F(k_0,\delta g)}{1+\xi^2 \delta k^2},
\end{equation}
where $A=(4\lambda_2-2\lambda_1)$ and $\delta g=(2\mu+2\lambda_1-2\lambda_2)$. 
This clearly indicate $\xi \propto |\delta g|^{-1}$, which implies the 
correlation length critical exponent $\nu=1$. The curvature function 
at the HSP $k_0=\pi$ can be written as 
$ F(k_0,\delta g)= \frac{2(2\lambda_2-\lambda_1)}{\delta g}$. 
As we approach the critical line $\lambda_2=\mu+\lambda_1$, 
curvature function is $ F(k_0,\delta g)\propto |\delta g|^{-1}$ 
which implies the value of $\gamma=1$. Thus we obtain a set 
of critical exponents i.e, $(z,\nu,\gamma)=(1,1,1)$ for the 
transition between gapped phases at $k_0=\pi$. Note that the 
critical exponents obey the scaling law in Eq.\ref{scaling_law}. 
Since the spectra on the critical line is linear around the gap 
closing point with $z=1$, the scaling law $\nu=\gamma$ is 
obeyed. Even though there is a transition between $w=1$ 
and $w=2$ gapped phases for both $k_0=0$ and $k_0=\pi$ 
HSPs, the nature of energy spectra, critical theory and the 
scaling of curvature function are different. This results in the 
modified scaling law observed previously for CP-2 at $k_0=0$. 

\subsubsection*{CRG for the transition between gapless phases}
\begin{figure}
	\includegraphics[width=6cm,height=6cm]{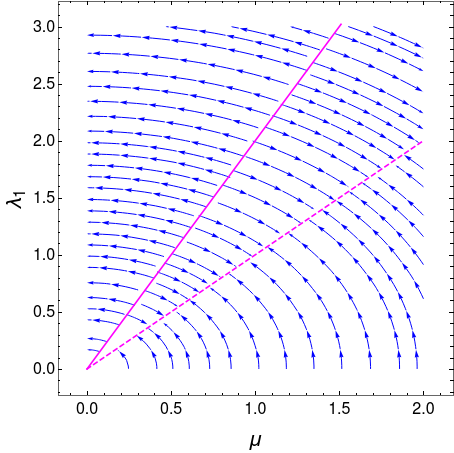}
	\caption{RG flow lines on the critical line $\lambda_2=\mu-\lambda_1$. 
		RG flow are away from the line $\lambda_1=2\mu$ (solid line) and towards 
		the line $\lambda_1=\mu$ (dashed line) which are critical and fixed lines
		respectively.}\label{RG_flow_gapless}
\end{figure}

In this section we discuss the topological transition between the 
gapless phases through multi-critical point on the critical line 
$\lambda_2=\mu-\lambda_1$. The gapless phases CP-1 and CP-2 
are found to have different set of critical exponents. The nature of transition 
between these two distinct gapless phases is indeed topological and 
occurs through the multi-critical point `b' (see Fig.\ref{phase_diag2}). 
We perform CRG again and derive 
RG equations and critical exponents to prove the existence of topological 
transition between gapless phases and also to characterize the critical behavior at the 
multi-critical point. \\
Curvature function on the critical line $\lambda_2=\mu-\lambda_1$ can 
be obtained as
\begin{equation}
\begin{split}
F(k,\mathbf{M})&= \dfrac{d\phi_k}{dk}\\
&= \dfrac{d}{dk}\left[ \tan ^{-1}\left(\frac{2 (\mu -\lambda_1) \sin (2 k) +2 \lambda_1 \sin (k)}{2 \mu -2 (\mu -\lambda_1) \cos (2 k) -2 \lambda_1 \cos (k)}\right)\right] \\
&= -\frac{\lambda_1 (\lambda_1-2 \mu )}{2 \left(\lambda_1^2-2 \lambda_1 \mu +2 \mu ^2 + 2 \mu (\mu -\lambda_1) \cos (k) \right)}-1,
\end{split}
\end{equation}
where $\mathbf{M}= \left\lbrace \mu, \lambda_1 \right\rbrace $. 
\begin{figure}
	\includegraphics[width=13cm,height=6cm]{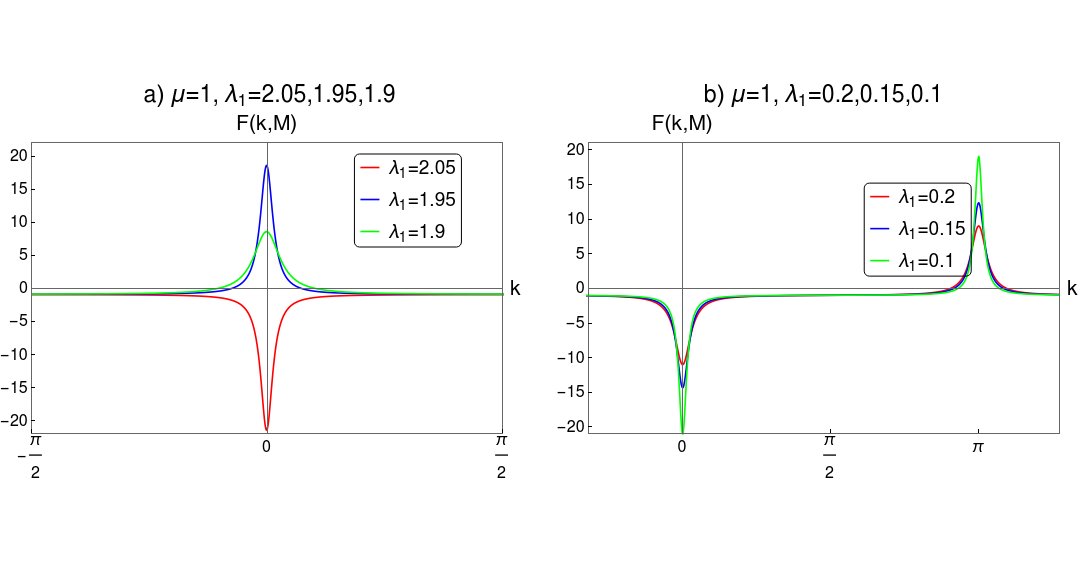}
	\caption{Curvature function $F(k,\mathbf{M})$ near the multi-critical points. (a) 
		Curvature function is plotted around the multi-critical point `b' in the phase diagram, which 
		distinguish between the distinct gapless phases (CP-1 and CP-2) on 
		the critical line $\lambda_2=\mu-\lambda_1$. (b) Curvature function is plotted around the multi-critical point `a' in the phase diagram. Both are plotted for several 
		values of $\lambda_1$ at $\mu=1$.} \label{curvature_plot_multi}
\end{figure}
Fig.\ref{curvature_plot_multi}(a) shows $F(k,\mathbf{M})$ for the transition
between gapless phases through multi-critical point. Surprisingly the 
curvature function tend to diverge as we approach the multi-critical 
point. For the parameter value $\mu=1$ multi-critical point is obtained 
at the critical value $\lambda_1=2$. Curvature function shows diverging peak as we approach critical value and flips 
sign across it. This behavior of the curvature function allow
one to perform CRG to understand the topological transition
between gapless phases. \\
The behavior of curvature function at the multi-critical point `a' is shown in
Fig.\ref{curvature_plot_multi}(b). It is a trivial multi-critical point at which
two critical line, $\lambda_2=\mu+\lambda_1$ and $\lambda_2=\mu-\lambda_1$ meet.
Hence, as we approach this multi-critical point from either directions 
the curvature function diverges at both HSPs $k_0=0$ and $k_0=\pi$. 
This multi-critical point preserve Lorentz invariance and no topological 
transition occurs between gapless phases as in the case of the 
multi-critical point `b'.\\
The RG flow 
equations, which signals the topological transition between the 
gapless phases through multi-critical point, for the coupling 
parameters $\lambda_1$ and $\mu$, can be derived as (refer to
`Method' section for a detailed derivation)
\begin{equation}
\frac{d\lambda_1}{dl}=-\frac{\lambda_1(\lambda_1-\mu)}{2(\lambda_1-2\mu)} \quad \text{and} \quad
\frac{d\mu}{dl}=-\frac{\mu(\mu-\lambda_1)}{2(\lambda_1-2\mu)}.
\end{equation}   
One can immediately spot a critical line for $\lambda_1=2\mu$ and 
a fixed line for $\lambda_1=\mu$ at which the RG equations satisfy 
the condition 
\begin{equation}
\left|\dfrac{d\lambda_1}{dl}\right| =\left|\dfrac{d\mu}{dl}\right| \rightarrow \infty \quad \text{and} \quad
\left|\dfrac{d\lambda_1}{dl}\right|=\left|\dfrac{d\mu}{dl}\right| \rightarrow 0
\end{equation}
The RG flow lines for the coupling parameters $\lambda_1$ and $\mu$ is shown in Fig.\ref{RG_flow_gapless}.  
Quantum
critical line and fixed line are represented as solid and dashed lines respectively. 
The line $\lambda_1=2\mu$, solid line in Fig.\ref{RG_flow_gapless}, 
indicate the multi-critical points 
for different values of $\mu$. This line distinguish between the $w=0$ 
(CP-1) and $w=1$ (CP-2) gapless phases 
on the critical line $\lambda_2=\mu-\lambda_1$. Therefore it indicate 
the TQPT between these gapless phases through the multi-critical point. 
The dashed line in Fig.\ref{RG_flow_gapless}, $\lambda_1=\mu$ represent 
fixed points in the flow diagram. The intersection of critical and fixed lines 
can be obtained analytically at $\mu=0$ and also can be 
observed at the same point in the flow diagram.\\
To characterize the critical behavior at the multi-critical point we calculate 
the critical exponents $(z,\nu,\gamma)$ as done in the case of gapped 
phases. Critical exponents can be calculated by expanding the Hamiltonian 
terms $\chi_z$ and $\chi_y$ from Eq.\ref{APS} on the critical line 
$\lambda_2=\mu-\lambda_1$, around the HSP $k_0=0$ upto third order.
\begin{align}
\chi_z &= \left( \frac{8\mu-6\lambda_1}{2}\right) \delta k^2 = B \delta k^2,
\end{align}
where $B= \left( \frac{8\mu-6\lambda_1}{2}\right)$, and 
\begin{align}
\chi_y&=-2(\lambda_1-2\mu)\delta k - \left( \frac{16\mu+ 18\lambda_1}{6}\right) \delta k^3 
= -2\delta g \delta k- A \delta k^3,
\end{align}
where $(\lambda_1-2\mu)=\delta g$ and $A=\left( \frac{16\mu+ 18\lambda_1}{6}\right) $. 
Now the Berry connection
can be written as (refer to `Method' section for details)
\begin{equation}
F(k,\delta g)= \frac{\left( \frac{-2B\delta g \delta k^2 + B A \delta k^4}{4 \delta g^2 \delta k^2}\right) }{1+\left( \frac{A^2+4\delta g B}{4 \delta g^2}\right) \delta k^2 + \left( \frac{B^2}{4 \delta g^2} \right) \delta k^4 }=\frac{F(k_0,\delta g)}{1+\xi^2\delta k^2+\xi^4 \delta k^4}.
\end{equation}
For different parameter values on the critical line, we observe the 
coefficient of $\delta k^4$ is dominant over $\delta k^2$. This implies 
the correlation length $\xi \propto |\delta g|^{-\frac{1}{2}}$, suggesting 
the correlation length exponent and dynamical critical exponents to 
be $\nu=\frac{1}{2}$ and $z=2$ respectively. To calculate the critical 
exponent $\gamma$ we obtain the curvature function at HSP, which 
has a form $F(k_0,\delta g) =\frac{4\mu-3\lambda_1}{2} |\delta g|^{-1}$. 
Therefore as we approach multi-critical point the curvature function critical 
exponent takes the value $\gamma=1$. Note that the scaling law is 
violated here also as in the case of the transition between the gapped 
phases $w=2$ and $w=1$ for $\lambda_2<0$. As proposed earlier the scaling law get 
modified as $\gamma=2\nu$ since the dynamical critical exponent $z=2$. 
Thus the critical phase at the multi-critical point, which governs the 
topological transition between two gapless phases on the critical 
line $\lambda_2=\mu-\lambda_1$, has critical exponents $(\nu,z,\gamma)=(\frac{1}{2},2,1)$. 

\subsubsection*{General discussions on RG flow behavior}
Here we discuss the general features of RG flow of coupling parameters for gapped phases. Behavior 
of RG flow lines are different for different quantum 
critical lines i.e, for $k_0=0$ and $k_0=\pi$, shown in Fig.\ref{CRG_flow_k0} and Fig.\ref{CRG_flow_kpi}. 
This difference is due to the distinct nature of 
fixed lines for both HSPs. In Fig.\ref{CRG_flow_k0} we observe
the fixed line at $\lambda_2=\frac{\lambda_1(\lambda_1+\mu)}{\lambda_1-8\mu}$. This fixed line is stable for finite range of parameter values
and flow lines flows towards it. However,
it is not the same case in Fig.\ref{CRG_flow_kpi}. The fixed 
line occurs at $\lambda_2=\frac{\lambda_1(\mu-\lambda_1)}{8\mu+\lambda_1} $, which  has both stable and unstable parts.
This causes a major distortion in the RG flow on $\lambda_1$-$\lambda_2$
plane. Thus the nature of RG flow are different for 
different critical lines.\\
\begin{figure}
	\includegraphics[width=13cm,height=7cm]{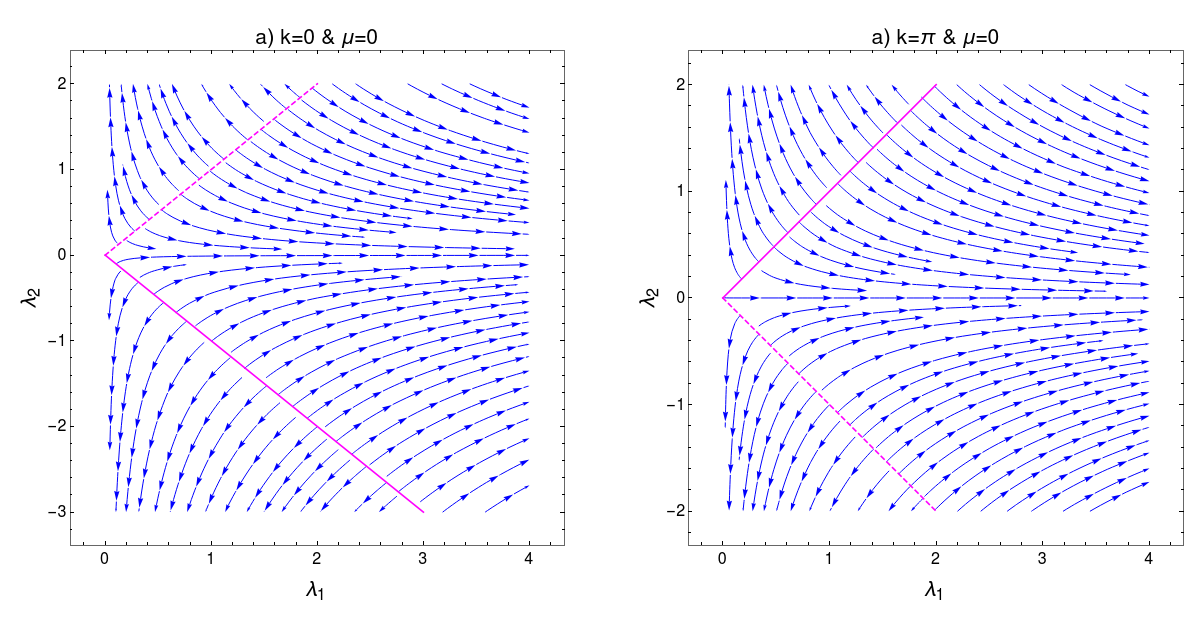}
	\caption{a) Flow diagram for $k_0=0$ (Eq.\ref{RG-lam1} and Eq.\ref{RG-lam2}) at $\mu=0$, b) Flow diagram for $k_0=\pi$ (Eq.\ref{RGklam1} and Eq.\ref{RGklam2}) at $\mu=0$. The RG flow directions are 
		pointed by the arrows. The critical lines are shown as solid lines and fixed 
		lines as dashed lines. } \label{CRG_flow_mu0}
\end{figure}
An interesting point can be observed when one set the parameter
$\mu=0$. RG flow in this case is shown in Fig.\ref{CRG_flow_mu0} for both
HSPs. Setting $\mu=0$, removes non-topological phase ($w=0$)
completely
and only topological gapped phases remain. It also eliminate the 
non-trivial multi-critical point along with distinct gapless phases.
Hence, the RG flow at $\mu=0$ for both HSPs are similar in nature.
The fixed lines for both HSPs are unstable with RG flow
lines flowing away. It is interesting to note that for $k_0=0$  (Fig.\ref{CRG_flow_mu0}.(a)),
the fixed line coincide with critical line for $k_0=\pi$. Similarly for $k_0=\pi$ (Fig.\ref{CRG_flow_mu0}.(b))
the fixed line coincide with the critical line for $k_0=0$. \\
RG flow lines in Fig.\ref{CRG_flow_k0} shows asymptotic nature 
around the the line $\lambda_1=2\mu$. The flow direction is 
reversed on the opposite sides of the multi-critical point,
which occurs at the intersection of fixed and critical lines.
This nature of RG flow lines are due to the term $\lambda_1-2\mu$ in
the denominator of RG equation for $\lambda_2$ in Eq.\ref{RG-lam2}.
This RG equation blows up for $\lambda_1=2\mu$ which accounts for the
asymptotic nature of RG flow lines in Fig.\ref{CRG_flow_k0}. For $\lambda_2$ value above the multi-critical point,
RG flow asymptotically increase for $\lambda_1<2\mu$ and asymptotically decrease for $\lambda_1>2\mu$. This flow directions
reverses for $\lambda_2$ value below the multi-critical point.
Similar nature can be expected for HSP $k_0=\pi$ around the line
$\lambda_1=-2\mu$.
\begin{figure}
	\includegraphics[width=12cm,height=12cm]{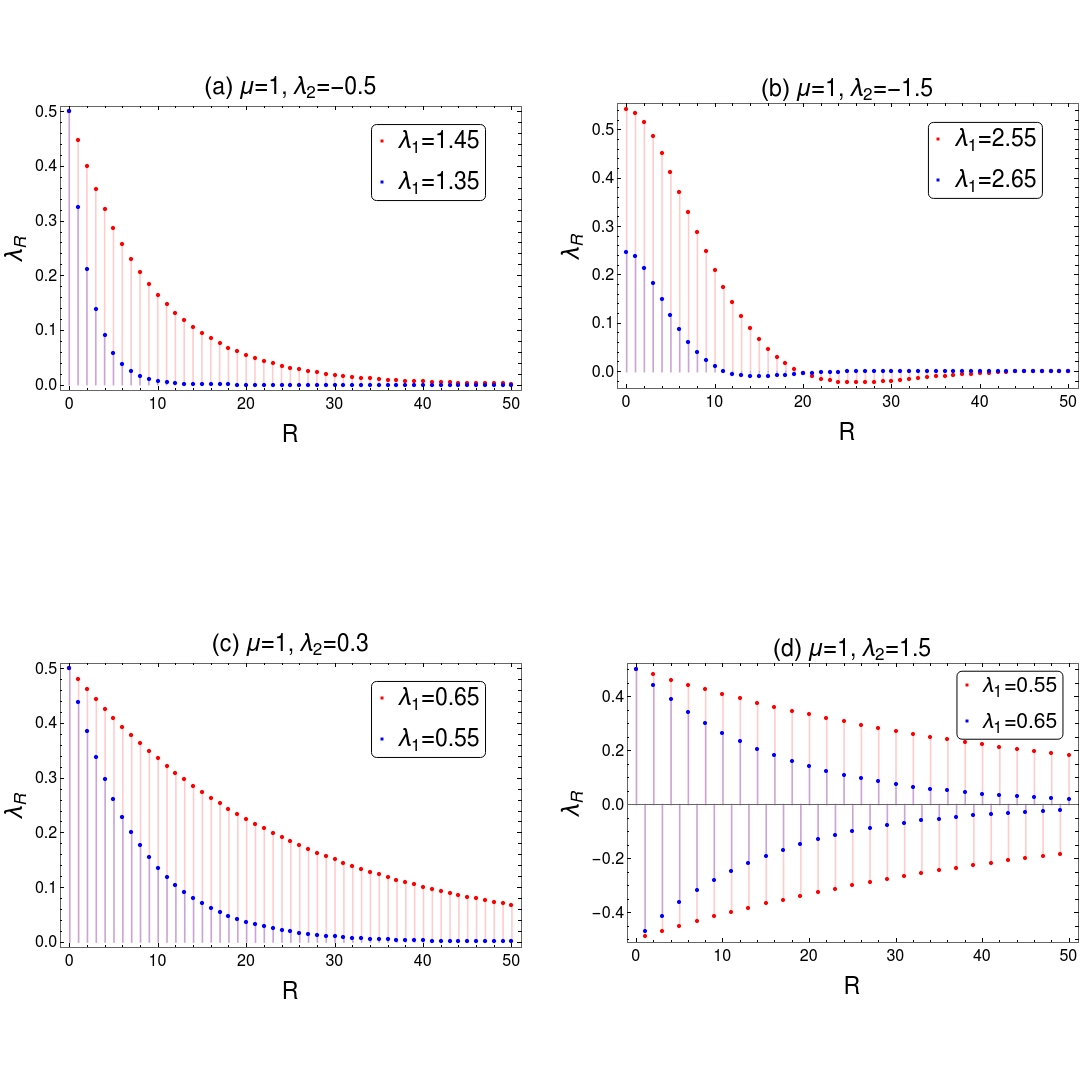}
	\caption{Behavior of correlation function $\lambda_R$ near quantum critical lines. (a) 
		$\lambda_R$ is plotted near the critical line $\lambda_2=\mu-\lambda_1$ for the transition 
		between $w=0$ and $w=1$ (i.e CP-1) with $\lambda_2<0$, where critical value of $\lambda_1=1.5$. (b) 
		$\lambda_R$ is plotted near the critical line $\lambda_2=\mu-\lambda_1$ for the transition 
		between $w=2$ and $w=1$ (i.e CP-2), where critical value of $\lambda_1=2.5$. (c) $\lambda_R$ 
		is plotted near the critical line $\lambda_2=\mu-\lambda_1$ for the transition between $w=0$ 
		and $w=1$ (i.e CP-1) with $\lambda_2>0$, where critical value of $\lambda_1=0.7$. (d) 
		$\lambda_R$ is plotted near the critical line $\lambda_2=\mu+\lambda_1$ for the transition 
		between $w=2$ and $w=1$, where the critical value of $\lambda_1=0.5$.} \label{correlation1}
\end{figure}
\begin{figure}
	\includegraphics[width=7cm,height=5cm]{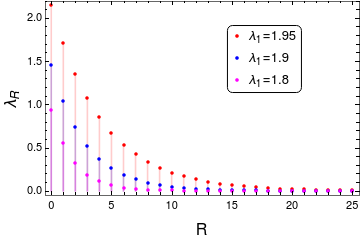}
	\caption{Behavior of correlation function $\lambda_R$ near the multi-critical point 
		$\lambda_1=2\mu$. The critical value $\lambda_1=2$ with $\mu=1$.}\label{correlation2}
\end{figure}
\subsubsection*{Correlation function for gapped and gapless phases }
Now we discuss the 
physical significance of correlation length as a length scale
to determine the correlation between Wannier states.
In the case of one dimensional systems, the curvature
function
is given by the Berry connection
$ F (k, \mathbf{M}) = 
\sum_{n} \left\langle u_{kn} | i \delta_k | u_{kn} \right\rangle  $, 
where $n$ is the index of all occupied bands. The Fourier transform of
which gives the charge polarization correlation function ($ \lambda_R $),
between Wannier states at a distance $R$ apart \cite{chen2017correlation,molignini2018universal}.\\
\begin{equation}
\lambda_R = \int \frac{dk}{2 \pi} e^{i k.R} F (k, \mathbf{M}) 
= \int \frac{dk}{2 \pi} e^{i k.R}  
\sum_{n} \left\langle u_{kn} | i \delta_k | u_{kn} \right\rangle  
=  
\sum_{n} \left\langle  Rn | r |0 n \right\rangle  . 
\end{equation}
We have two bands in our model and only the lower
band $(n=1)$ is occupied. 
Therefore we have 
$\lambda_R = \left\langle R|r|0\right\rangle  $, 
which is a measure of overlap between Wannier
states at $0$ and $R$. The zeroth component $\lambda_0$ is the
charge polarization, which is the topological invariant.
Since Wannier state $ \left\langle r|R\right\rangle  = W (r - R) $ is a localized
function with center at $R$, the quantity $ \left\langle R|r|0\right\rangle $ is 
expected to decay with $R$ to zero.\\
The correlation function $\lambda_R$ can be analytically calculated in the 
continuous approximation for the appropriate gauge choice of Berry connection, which takes
Ornstein-Zernike form. We study the behavior of $\lambda_R$ near the critical line 
$\lambda_2=\mu-\lambda_1$ which occurs at the HSP $k_0=0$. Since the critical line has 
distinct gapless phases (CP-1 and CP-2), we study the nature of $\lambda_R$ separately 
near these gapless phases.  As we approach the CP-1 i.e, for the transition between gapped 
$w=0$ to $w=1$ phase, the correlation function $\lambda_R$ can be obtained as (refer to `Method' section for details)
\begin{equation}
\lambda_R=\frac{1}{2\xi} \left( \frac{2 (\lambda_1+2 \lambda_2)}{2 \mu-2 \lambda_1-2 \lambda_2}\right)\exp \left(-\frac{|R|}{\xi}\right)
\label{corrCP1},
\end{equation}
where $\xi=\frac{2 (\lambda_1+2 \lambda_2)}{2 \mu-2 \lambda_1-2 \lambda_2}$.
Similarly as we approach the CP-2 i.e, for the transition between gapped $w=1$ to $w=2$ phase, $\lambda_R$ can be obtained as (refer to `Method' section for details)
\begin{equation}
\lambda_R=\frac{1}{2\;\xi\sqrt{2}}\left( \frac{2 (\lambda_1+2 \lambda_2)}{2 \mu-2 \lambda_1-2 \lambda_2}\right)
\left\lbrace  \sin \left(\frac{\left| R\right| }{\sqrt{2}\;\xi}\right)+\cos \left(\frac{\left| R\right| }{\sqrt{2}\;\xi}\right)\right\rbrace  \exp{\left(- \frac{|R|}{\sqrt{2} \;\xi}\right) }
\label{corrCP2},
\end{equation}
where $\xi= \sqrt{\frac{2 \lambda_1+8 \lambda_2}{2 (2 \mu-2\lambda_1-2\lambda_2)}}$.
Behavior of correlation function near the critical lines between distinct gapped phases is 
depicted in Fig.\ref{correlation1}.
Fig.\ref{correlation1}(a) shows the decay in the correlation function in Eq.\ref{corrCP1} 
as we approach a critical point at $\lambda_1=1.5$ on CP-1. We observe the decay length of the $\lambda_R$ is shorter
for the parameter value away from the critical value and it gets longer as we approach 
the critical point. In other words the correlation function decays slower 
near the critical line as the decay is sharp deep inside the gapped phase.
Similar behavior can be observed for the transition across CP-2 as shown in
Fig.\ref{correlation1}(b). In this case the critical point is at $\lambda_1=2.5$.
$\lambda_R$ shows sharp decay for the parameter value away from the critical value and 
the decay length is longer as we approach the critical point. This indicate the 
TQPT between the gapped phases as this behavior of correlation function is universal 
around a QCP. Note that for $w= 0$ gapped phase, $\lambda_2$ range from $-1$ to $1$ (see Fig.\ref{phase_diag2}). 
In this range of $\lambda_2 $,
we consider one attractive (-ve) and the other one repulsive (+ve) coupling. Fig.\ref{correlation1}(a) 
is plotted for attractive coupling of $\lambda_2$ and Fig.\ref{correlation1}(c) is plotted 
for repulsive coupling of $\lambda_2$. The critical value of $\lambda_1=0.7$ near to 
which $\lambda_R$ decay slowly and sharp decay can be observed for the value away 
from critical value. We observe the decay in $\lambda_R$ is much slower in the repulsive 
case than in the attractive case at the same distance from the critical line.\\ 
The topological transition across the critical line $\lambda_2=\mu+\lambda_1$ can also 
be observed in terms $\lambda_R$. This critical line corresponds to the transition between 
gapped phases with $w=2$ and $w=1$.
Behavior of $\lambda_R$ for the HSP $k_0=\pi$ can be obtained as (refer to `Method' section for details)
\begin{equation}
\lambda_R=\frac{(-1)^R}{2\;\xi}\left( \frac{2 (2 \lambda_2 -\lambda_1)}{2 \lambda_1-2 \lambda_2 +2 \mu }\right)  \exp \left(-\frac{|R|}{\xi}\right),
\end{equation}
where $\xi=\left( \frac{4 \lambda_2-2 \lambda_1}{2 \mu+2 \lambda_1 -2 \lambda_2 }\right)$.
Fig.\ref{correlation1}(d) shows oscillatory behavior of $\lambda_R$ close to the critical point at $\lambda_1=0.5$ on the critical line $\lambda_2=\mu+\lambda_1$. We observe that the amplitude of the oscillation decreases, which indicate the decay in $\lambda_R$. This decay gets slower as we approach the critical point as in Fig.\ref{correlation1}(d). This clearly confirms the presence of TQPT across the critical point between the gapped phases $w=1$ and $w=2$. \\
Behavior of correlation function $\lambda_R$ near a critical point signals the TQPT successfully. Therefore we analyze the same universal property of $\lambda_R$ for the transition between gapless phases CP-1 and CP-2.   
The analytical expression for the gapless excitation of 
the correlation function $\lambda_R$ can be obtained as (refer to `Method' section for details)
\begin{equation}
\lambda_R=\frac{1}{2\;\xi\sqrt{2}}\left( \frac{4 \mu -3 \lambda_1}{2 (\lambda_1-2 \mu )}\right)
\left\lbrace  \sin \left(\frac{\left| R\right| }{\sqrt{2}\;\xi}\right)+\cos \left(\frac{\left| R\right| }{\sqrt{2}\;\xi}\right)\right\rbrace  \exp{\left(-\frac{|R|}{\xi}\right) },
\end{equation}
where $\xi= \sqrt{\frac{8 \mu-6 \lambda_1}{4 (\lambda_1-2\mu)}}$.
Fig.\ref{correlation2} shows the behavior of $\lambda_R$ as we approach the multi-critical point at $\lambda_1=2$. $\lambda_R$ decays sharply deep within the gapless phase and the decay length increases as the $\lambda_1$ value approaches critical point. The decay tends to slow down with longer decay length for the value close to critical point. This behavior of $\lambda_R$ near the multi-critical point is similar to the cases of gapped phases. One can conclude from the behavior of $\lambda_R$ in Fig.\ref{correlation2} that it clearly indicate the presence of TQPT across the multi-critical point between the gapless phases CP-1 and CP-2. 

\section*{Discussion}
The theory 
of critical phenomena and curvature function renormalization scheme, developed for 
the topological phase transitions, provides an alternative platform to understand the 
transition between gapped phases against the conventional theory 
on topological invariant. We have shown explicitly that
these tools can also be extended for the characterization of topological quantum
phase transition 
occurring between gapless phases. The two distinct gapless phases of our 
model Hamiltonian has been analyzed and they
were found to belong to different universality classes based 
on the values of critical exponents.
Among the three quantum critical lines of the model Hamiltonian, 
	two are topological in nature
	and also capture the essential TQPT across the gapless topological quantum critical line. This interesting 
	feature is absent in the original Kitaev chain.
CRG analysis confirmed the 
presence of topological quantum phase transition between the gapless phases
through the non-trivial multi-critical 
point. 
We have shown explicitly the break
down of Lorentz invariance at the topological multi-critical point.
The values of critical exponents 
revealed that the transition is in the Lifshitz universality class.  
We have performed the calculation of Wannier state correlation function
for the TQPT between gapped and gapless phases. Decrease in the decay rate of correlation 
function as we approach multi-critical point revealed the presence of TQPT between gapless phases.

\section*{Methods}

\subsection*{Derivation of CRG equations}
\textbf{For gapped phases:}
Here, we derive the RG equations for $k_0=0$.  
Referring the generic form of the RG equation in Eq.\ref{CRG_eq} we 
obtain three RG equations corresponding to the parameters. 
Curvature function can be obtained as
\begin{equation}
F(k,\mathbf{M})= \frac{\lambda_1  \cos (k) (\mu -3 \lambda_2)+2 \lambda_2 \mu  \cos (2 k)-\lambda_1^2-2 \lambda_2^2}{2 \lambda_1 \cos (k) (\lambda_2-\mu )-2 \lambda_2 \mu  \cos (2 k)+\lambda_1^2+\lambda_2^2+\mu ^2},
\end{equation}
where $\mathbf{M}= \left\lbrace \lambda_1,\lambda_2,\mu\right\rbrace  $. Second derivative of $F(k,\mathbf{M})$ at $k_0=0$ is
\begin{equation}
\partial_k^2 F(k,\mathbf{M})|_{k=0}=\frac{(\text{$\lambda_2$}+\mu ) \left(\text{$\lambda_1$}^2+\text{$\lambda_1$} (\mu -\text{$\lambda_2$})+8 \text{$\lambda_2 $} \mu \right)}{(\text{$\lambda_1 $}+\text{$\lambda_2 $}-\mu )^3}.
\end{equation}
Derivative of the curvature function at $k_0=0$ with respect to
the parameters $\lambda_1,\lambda_2$ 
and $\mu$ are correspondingly
\begin{align}
\partial_{\lambda_1}F(0,\mathbf{M}) &= \frac{\text{$\lambda_2 $}+\mu }{(\text{$\lambda_1 $}+\text{$\lambda_2 $}-\mu )^2},\\
\partial_{\lambda_2}F(0,\mathbf{M}) &= \frac{2 \mu -\text{$\lambda_1 $}}{(\text{$\lambda_1 $}+\text{$\lambda_2 $}-\mu )^2},\\
\partial_{\mu}F(0,\mathbf{M}) &= -\frac{\text{$\lambda_1 $}+2 \text{$\lambda_2 $}}{(\text{$\lambda_1 $}+\text{$\lambda_2 $}-\mu )^2}.
\end{align} 
This gives three RG equations for the parameters as
\begin{equation}
\begin{split}
\frac{d\lambda_1}{dl}&= \frac{1}{2} \frac{(\text{$\lambda_2 $}+\mu ) \left(\text{$\lambda_1 $}^2+\text{$\lambda_1 $} (\mu -\text{$\lambda_1 $})+8 \text{$\lambda_1 $} \mu \right) (\text{$\lambda_1 $}+\text{$\lambda_2 $}-\mu )^2}{(\text{$\lambda_1 $}+\text{$\lambda_2 $}-\mu )^3 (\text{$\lambda_2 $}+\mu)}\\ &=\frac{\text{$\lambda_1 $}^2+\text{$\lambda_1$} (\mu -\text{$\lambda_2 $})+8 \text{$\lambda_2 $} \mu }{2 (\text{$\lambda_1 $}+\text{$\lambda_2 $}-\mu )},
\end{split}
\end{equation}
\begin{equation}
\begin{split}
\frac{d\lambda_2}{dl}&=\frac{1}{2}\frac{(\text{$\lambda_2$}+\mu ) \left(\text{$\lambda_1 $}^2+\text{$\lambda_1 $} (\mu -\text{$\lambda_2 $})+8 \text{$\lambda_2 $} \mu \right) (\text{$\lambda_1 $}+\text{$\lambda_2 $}-\mu )^2}{(\text{$\lambda_1 $}+\text{$\lambda_2 $}-\mu )^3 (2 \mu -\text{$\lambda_1 $})}\\
&=-\frac{(\text{$\lambda_2 $}+\mu ) \left(\text{$\lambda_1 $}^2+\text{$\lambda_1 $} (\mu -\text{$\lambda_2 $})+8 \text{$\lambda_2 $} \mu \right)}{2 (\text{$\lambda_1 $}-2 \mu ) (\text{$\lambda_1 $}+\text{$\lambda_2 $}-\mu )},
\end{split}
\end{equation}
\begin{equation}
\begin{split}
\frac{d\mu}{dl}&=-\frac{1}{2} \frac{(\text{$\lambda_2 $}+\mu ) \left(\text{$\lambda_1 $}^2+\text{$\lambda_1 $} (\mu -\text{$\lambda_2 $})+8 \text{$\lambda_2 $} \mu \right) (\text{$\lambda_1$}+\text{$\lambda_2 $}-\mu )^2}{(\text{$\lambda_1 $}+\text{$\lambda_2 $}-\mu )^3 (\text{$\lambda_1 $}+2 \text{$\lambda_2 $})}\\
&=-\frac{(\text{$\lambda_2 $}+\mu ) \left(\text{$\lambda_1 $}^2+\text{$\lambda_1 $} (\mu -\text{$\lambda_2 $})+8 \text{$\lambda_2 $} \mu \right)}{2 (\text{$\lambda_1 $}+2 \text{$\lambda_2 $}) (\text{$\lambda_1 $}+\text{$\lambda_2 $}-\mu )}.
\end{split}
\end{equation}
Following the similarly procedure one can obtain RG equations for HSP $k_0=\pi$. Second derivative of $F(k,\mathbf{M})$ is taken at $k_0=\pi$
\begin{equation}
\partial_k^2 F(k,\mathbf{M})|_{k=\pi}=-\frac{(\text{$\lambda_2$}+\mu ) \left(\text{$\lambda_1$}^2+\text{$\lambda_1$} (\text{$\lambda_2$}-\mu)+8 \text{$\lambda_2 $} \mu \right)}{(\text{$\lambda_1 $}-\text{$\lambda_2 $}+\mu )^3}.
\end{equation}
Derivative of $F(k,\mathbf{M})$ at $k_0=\pi$ with respect to the 
parameters are
\begin{align}
\partial_{\lambda_1}F(\pi,\mathbf{M}) &= -\frac{\text{$\lambda_2 $}+\mu }{(\text{$\lambda_1 $}-\text{$\lambda_2 $}+\mu )^2},\\
\partial_{\lambda_2}F(\pi,\mathbf{M}) &= \frac{2 \mu +\text{$\lambda_1 $}}{(\text{$\lambda_1 $}-\text{$\lambda_2 $}+\mu )^2},\\
\partial_{\mu}F(\pi,\mathbf{M}) &= \frac{\text{$\lambda_1 $}-2 \text{$\lambda_2 $}}{(\text{$\lambda_1 $}-\text{$\lambda_2 $}+\mu )^2}.
\end{align} 
After few steps of calculation one can arrive at the RG equations 
\begin{equation}
\frac{d\lambda_1}{dl}=\frac{\text{$\lambda_1 $}^2+\text{$\lambda_1$} (\text{$\lambda_2 $}-\mu)+8 \text{$\lambda_2 $} \mu }{2 (\text{$\lambda_1 $}-\text{$\lambda_2 $}+\mu )},
\end{equation}
\begin{equation}
\frac{d\lambda_2}{dl}=-\frac{(\text{$\lambda_2 $}+\mu ) \left(\text{$\lambda_1 $}^2+\text{$\lambda_1 $} (\text{$\lambda_2 $}-\mu)+8 \text{$\lambda_2 $} \mu \right)}{2 (\text{$\lambda_1 $}+2 \mu ) (\text{$\lambda_1 $}-\text{$\lambda_2 $}+\mu )},
\end{equation}
\begin{equation}
\frac{d\mu}{dl}=-\frac{(\text{$\lambda_2 $}+\mu ) \left(\text{$\lambda_1 $}^2+\text{$\lambda_1 $} (\text{$\lambda_2 $}-\mu)+8 \text{$\lambda_2 $} \mu \right)}{2 (\text{$\lambda_1 $}-2 \text{$\lambda_2 $}) (\text{$\lambda_1 $}-\text{$\lambda_2 $}+\mu )}.
\end{equation}\\

\noindent\textbf{For gapless phases:}
As in the case of gapped phases, CRG can be performed for gapless 
phases as well. In our model, curvature function on the critical line 
$\lambda_2=\mu-\lambda_1$ is
\begin{equation}
F(k,\mathbf{M})= -\frac{\lambda_1 (\lambda_1-2 \mu )}{2 \left(\lambda_1^2-2 \lambda_1 \mu +2 \mu ^2 + 2 \mu (\mu -\lambda_1) \cos (k) \right)}-1,
\end{equation}
here $\mathbf{M}=\left\lbrace \lambda_1,\mu\right\rbrace $. Second derivative of curvature 
function at $k_0=0$ can be obtained as
\begin{equation}
\partial_k^2 F(k,\mathbf{M})|_{k=0}= \frac{\text{$\lambda_1$} \mu  (\text{$\lambda_1 $}-2 \mu ) (\text{$\lambda_1 $}-\mu ) \left(\text{$\lambda_1 $}^2-2 \text{$\lambda_1 $} \mu -2 \mu  (\text{$\lambda_1 $}-\mu )+2 \mu ^2\right)}{\left(\text{$\lambda_1 $}^2-2 \text{$\lambda_1 $} \mu +2 \mu  (\mu -\text{$\lambda_1 $})+2 \mu ^2\right)^3}.
\end{equation}
Derivative of curvature function with respect to the parameters $\lambda_1$ 
and $\mu$ are correspondingly
\begin{align}
\partial_{\lambda_1}F(0,\mathbf{M}) &= \frac{\mu }{(\text{$\lambda_1 $}-2 \mu )^2},\\
\partial_{\mu}F(0,\mathbf{M}) &= -\frac{\text{$\lambda_1 $}}{(\text{$\lambda_1 $}-2 \mu )^2}.
\end{align} 
This gives RG equations for the parameters as
\begin{equation}
\begin{split}
\frac{d\lambda_1}{dl}&= \frac{1}{2}\frac{\text{$\lambda_1$} \mu  (\text{$\lambda_1 $}-2 \mu ) (\text{$\lambda_1 $}-\mu ) \left(\text{$\lambda_1 $}^2-2 \text{$\lambda_1 $} \mu -2 \mu  (\text{$\lambda_1 $}-\mu )+2 \mu ^2\right) (\text{$\lambda_1 $}-2 \mu )^2}{\mu\left(\text{$\lambda_1 $}^2-2 \text{$\lambda_1 $} \mu +2 \mu  (\mu -\text{$\lambda_1 $})+2 \mu ^2\right)^3}\\ 
&=-\frac{\text{$\lambda_1$} (\text{$\lambda_1 $}-\mu )}{2 (\text{$\lambda_1 $}-2 \mu )},
\end{split}
\end{equation}
\begin{equation}
\begin{split}
\frac{d\mu}{dl}&= -\frac{1}{2}\frac{\text{$\lambda_1$} \mu  (\text{$\lambda_1 $}-2 \mu ) (\text{$\lambda_1 $}-\mu ) \left(\text{$\lambda_1 $}^2-2 \text{$\lambda_1 $} \mu -2 \mu  (\text{$\lambda_1 $}-\mu )+2 \mu ^2\right) (\text{$\lambda_1 $}-2 \mu )^2}{\lambda_1\left(\text{$\lambda_1 $}^2-2 \text{$\lambda_1 $} \mu +2 \mu  (\mu -\text{$\lambda_1 $})+2 \mu ^2\right)^3}\\ 
&=-\frac{\mu  (\mu -\text{$\lambda_1$})}{2 (\text{$\lambda_1 $}-2 \mu )}.
\end{split}
\end{equation}
\subsection*{Derivation of critical exponents}
\textbf{For gapped phases:}
Components of the Hamiltonian, $ \chi_{z} (k) = -2 \lambda_1 \cos k - 2 \lambda_2 \cos 2k 
+ 2\mu,$ and $ \chi_{y} (k) = 2 \lambda_1 \sin k + 2 \lambda_2 \sin 2k$, 
are expanded around HSP $k_0=0$ as \\
\begin{align}
\chi_z&= (2\mu-2\lambda_1-2\lambda_2)+\frac{(8\lambda_2+2\lambda_1)}{2} \delta k^2=\delta g + B \delta k^2,\\
\chi_y&= (4\lambda_2+2\lambda_1)\delta k =A \delta k
\end{align}
We perform the expansion of $\chi_{z} (k)$ and $ \chi_{y} (k)$ for  
HSP $k_0=\pi$ only upto first order, since the higher order terms are 
insignificant due to linear spectra around $k_0=\pi$. Thus we have
\begin{align}
\chi_z&= (2\mu+2\lambda_1-2\lambda_2)=\delta g ,\\
\chi_y&= (4\lambda_2-2\lambda_1)\delta k=A \delta k
\end{align}
Curvature function for 1D systems can be written in terms of $\chi_{z} (k)$ and $ \chi_{y} (k)$ as
\begin{equation}
F(k,\mathbf{M})= \frac{\chi_y\partial_k \chi_z - \chi_z \partial_k \chi_y}{\chi_z^2+\chi_y^2}.
\end{equation}
In the vicinity of HSPs one can write the curvature function in 
Ornstein-Zernike form in Eq.\ref{Lorenzian}. For HSP $k_0=0$ it reads 
\begin{equation}
\begin{split}
F(k,\delta g)&= \frac{\left( A\delta k (2B\delta k)-(\delta g+B\delta k^2)A\right) }{\delta g^2 +(2 \delta g B+A^2) \delta k^2 + B^2\delta k^4 }\\
&= \frac{\left( \frac{2BA\delta k^2-A(\delta g+B\delta k^2)}{\delta g^2}\right) }{1 + \frac{(2 \delta g B+A^2)}{\delta g^2} \delta k^2 + \frac{B^2}{\delta g^2}\delta k^4 }\\
&= \frac{F(k_0,\delta g)}{1+\xi^2 \delta k^2+\xi^4\delta k^4},
\end{split}
\end{equation}
where $F(k_0,\delta g)=\frac{2(\lambda_1+2\lambda_2)}{(2\mu-2\lambda_1-2\lambda_2)} \propto |\delta g|^{-1} \implies \gamma=1$. Correlation length $\xi$ for the transition 
between $w=0$ and $w=1$ gapped phases is $\xi=\frac{(4\lambda_2+2\lambda_1)}{(2\mu-2\lambda_1-2\lambda_2)}\propto |\delta g|^{-1} \implies \nu=1$, since $\delta k^2$ term dominates over $\delta k^4$. Similarly for 
the transition between $w=2$ and $w=1$ gapped phases  $\xi=\sqrt{\frac{(8\lambda_2+2\lambda_1)}{2(2\mu-2\lambda_1-2\lambda_2)}}\propto |\delta g|^{-\frac{1}{2}} \implies \nu=\frac{1}{2}$, since $\delta k^4$ term dominates 
over $\delta k^2$.\\
Following the same procedure in the vicinity of HSP $k_0=\pi$, 
the curvature function can be written as
\begin{equation}
\begin{split}
F(k,\delta g)&= \frac{\left( \frac{A}{\delta g}\right) }{1 + \frac{(A^2)}{\delta g^2} \delta k^2}\\
&= \frac{F(k_0,\delta g)}{1+\xi^2 \delta k^2},
\end{split}
\end{equation}
where $F(k_0,\delta g)=\frac{2(2\lambda_2-\lambda_1)}{(2\mu+2\lambda_1-2\lambda_2)} \propto |\delta g|^{-1} \implies \gamma=1$. The correlation length  $\xi=\frac{(4\lambda_2-2\lambda_1)}{(2\mu+2\lambda_1-2\lambda_2)}\propto |\delta g|^{-1} \implies \nu=1$.\\\\
\noindent\textbf{For gapless phases:}
Components of the Hamiltonian expanded around the HSP $k_0=0$, on the critical line $\lambda_2=\mu-\lambda_1$ are,
\begin{align}
\chi_z&= 2\mu-2\lambda_1\cos k-2\mu \cos 2k + 2\lambda_1 \cos 2k\\
&= \left( \frac{8\mu-6\lambda_1}{2}\right) \delta k^2 = B \delta k^2,
\end{align}
\begin{align}
\chi_y&= 2\lambda_1\sin k-2\mu \sin 2k + 2\lambda_1 \sin 2k\\
&=-2(\lambda_1-2\mu)\delta k - \left( \frac{16\mu+ 18\lambda_1}{6}\right) \delta k^3 = -2\delta g \delta k- A \delta k^3,
\end{align}
where $(\lambda_1-2\mu)=\delta g$. The curvature function in 
Ornstein-Zernike form in Eq.\ref{Lorenzian}, can be written as
\begin{align}
F(k,\delta g)&=\frac{(-2\delta g \delta k-A\delta k^3)2B\delta k-B\delta k^2(-2\delta t- 3A \delta k^2)}{(B\delta k^2)^2 +(-2\delta g \delta k-A\delta k^3)^2}\\
&= \frac{-2B\delta g \delta k^2 + AB \delta k^4}{4 \delta g^2 \delta k^2 + (B^2+4\delta g A) \delta k^4 + A^2 \delta k^6}\\
&= \frac{\left( \frac{-2B\delta g \delta k^2 + B A \delta k^4}{4 \delta g^2 \delta k^2}\right) }{1+\left( \frac{A^2+4\delta g B}{4 \delta g^2}\right) \delta k^2 + \left( \frac{B^2}{4 \delta g^2} \right) \delta k^4 }\\
&=\frac{F(k_0,\delta g)}{1+\xi^2\delta k^2+\xi^4 \delta k^4},
\end{align}
where $F(k_0,\delta g)=\frac{(4\mu-3\lambda_1)}{2(\lambda_1-2\mu)} \propto |\delta g|^{-1} \implies \gamma=1$. The correlation length $\xi=\sqrt{\frac{8\mu-6\lambda_1}{4(\lambda_1-2\mu)}}\propto |\delta g|^{-\frac{1}{2}} \implies \nu=\frac{1}{2}$, since $\delta k^4$ term is dominant.

\subsection*{Derivation of modified scaling law}
In order to preserve the constant value of topological invariant, 
the divergence of the curvature function near HSP, as we approach 
the transition point ($\mathbf{M}\rightarrow\mathbf{M_c}$), has to 
be conserved \cite{luo2019advanced}. The contribution to the 
topological invariant from the divergence $C_{div}$ of curvature 
function near the HSP $k_0=0$, as we approach CP-2, can be 
obtained by integrating over the width $\xi^{-1}$
\begin{equation}
C_{div} =F(k_0,\delta g) \int\limits_{-\xi^{-1}}^{\xi^{-1}} \frac{d\delta k}{(1+\xi^4\delta k^4)},
\end{equation}
here
\begin{align}
\int\limits_{-\xi^{-1}}^{\xi^{-1}} \frac{d\delta k}{(1+\xi^4\delta k^4)} &= \frac{1}{\sqrt{\xi^4}}\tan^{-1}\left(\sqrt{\xi^4} \delta k^2 \right)|_{-\xi^{-1}}^{\xi^{-1}} \\
&= \frac{1}{\xi^2}\left(\tan^{-1}(-1) - \tan^{-1}(1)\right) \\
&=\frac{1}{\xi^2} \left( \frac{\pi}{2}\right) 
\end{align}
Thus we have 
\begin{equation}
C_{div} =\frac{F(k_0,\delta g)}{\xi^2} \times \mathbf{O}(1) = \text{constant}.
\end{equation}
Combining this with Eq.\ref{critical-exponents} (i.e, $F(k_0,\mathbf{M}) \propto |\mathbf{M}-\mathbf{M}_c|^{-\gamma} ,\;\;\; \xi \propto |\mathbf{M}-\mathbf{M}_c|^{-\nu}$), 
we get the modified scaling law for 1D as
\begin{equation}
\gamma=2\nu.
\end{equation}

\subsection*{Calculations of correlation function}
The critical line $\lambda_2=\mu-\lambda_1$ which occurs at $k_0=0$, has distinct gapless phases, CP-1 and CP-2.
As we approach the CP-1, the correlation function $\lambda_R$ can be obtained as
\begin{align}
\lambda_R &= \int_{-\infty}^{\infty} \frac{dk}{2\pi} e^{ikR}F(k,\mathbf{M}) \\
&= \int_{-\infty}^{\infty} \frac{dk}{2\pi} \frac{F(0,\mathbf{M})}{1+\xi^2k^2} e^{ikR}\\
&= \frac{F(0,\mathbf{M})}{2\xi} e^{-|R|/\xi}
\end{align}
In terms of the parameters of the model Hamiltonian the above equation reads 
\begin{equation}
\lambda_R=\frac{1}{2\xi} \left( \frac{2 (\lambda_1+2 \lambda_2)}{2 \mu-2 \lambda_1-2 \lambda_2}\right)\exp \left(-\frac{|R|}{\xi}\right)
\end{equation}
where $\xi=\frac{2 (\lambda_1+2 \lambda_2)}{2 \mu-2 \lambda_1-2 \lambda_2}$.
Similarly as we approach the CP-2, $\lambda_R$ can be obtained as
\begin{align}
\lambda_R &= \int_{-\infty}^{\infty} \frac{dk}{2\pi} e^{ikR}F(k,\mathbf{M}) \\
&= \int_{-\infty}^{\infty} \frac{dk}{2\pi} \frac{F(0,\mathbf{M})}{1+\xi^4k^4} e^{ikR}\\
&= \frac{F(0,\mathbf{M})}{2\sqrt{2}\xi} \left( \cos\left[ \frac{|R|}{\sqrt{2}\xi}\right] +\sin\left[ \frac{|R|}{\sqrt{2}\xi}\right] \right) e^{-|R|/\sqrt{2}\xi}
\end{align}
In terms of parameters of the model Hamiltonian it reads
\begin{equation}
\lambda_R=\frac{1}{2\;\xi\sqrt{2}}\left( \frac{2 (\lambda_1+2 \lambda_2)}{2 \mu-2 \lambda_1-2 \lambda_2}\right)
\left\lbrace  \sin \left(\frac{\left| R\right| }{\sqrt{2}\;\xi}\right)+\cos \left(\frac{\left| R\right| }{\sqrt{2}\;\xi}\right)\right\rbrace  \exp{\left(- \frac{|R|}{\sqrt{2} \;\xi}\right) }
\end{equation}
where $\xi= \sqrt{\frac{2 \lambda_1+8 \lambda_2}{2 (2 \mu-2\lambda_1-2\lambda_2)}}$.
For the critical line  $\lambda_2=\mu+\lambda_1$ which occurs at $k_0=\pi$,
the $\lambda_R$ can be obtained as
\begin{align}
\lambda_R &= \int_{-\infty}^{\infty} \frac{dk}{2\pi} e^{ikR}F(k,\mathbf{M}) \\
&= \int_{-\infty}^{\infty} \frac{dk}{2\pi} \frac{F(\pi,\mathbf{M})}{1+\xi^2k^2} e^{i(\pi+k)R}\\
&= F(\pi,\mathbf{M})\frac{e^{\left( i\pi R-\frac{R}{\xi}\right) }}{2\xi} \\
&= (-1)^R\frac{F(\pi,\mathbf{M})}{2\xi} e^{-|R|/\xi}
\end{align}
Since the bulk gap closes at $k_0=\pi$ the sign alternates between even and odd sites. In terms of the parameters of the model Hamiltonian the above equation reads 
\begin{equation}
\lambda_R=\frac{(-1)^R}{2\;\xi}\left( \frac{2 (2 \lambda_2 -\lambda_1)}{2 \lambda_1-2 \lambda_2 +2 \mu }\right)  \exp \left(-\frac{|R|}{\xi}\right),
\end{equation}
where $\xi=\left( \frac{4 \lambda_2-2 \lambda_1}{2 \mu+2 \lambda_1 -2 \lambda_2 }\right)$.

\section*{Acknowledgements}
S.S. would like to acknowledge DST (EMR/2017/000898) for 
the support. Authors would 
like to acknowledge ICTS for a useful discussion meeting on 
``Novel Phases of Quantum Matter". Authors would 
like to acknowledge Prof. Subir Sachdev, Prof. Diptiman Sen, Prof. Sumathi Rao for the useful discussions.
Authors would 
like to acknowledge Prof. R. Srikanth for reading the manuscript critically. R.R.K, Y.R.K. and S.R. would like to acknowledge 
PPISR, RRI library for the books and journals.

\section*{Author contributions statement}

S.S. identified the problem,  R.R.K. solved the problem and wrote the manuscript, Y.R.K and S.R conducted the numerical calculations. All authors analyzed the results and reviewed the manuscript. 

\section*{Additional information}
All authors declares no competing interests.

\end{document}